\documentclass[amsthm,3p, authoryear, 12pt]{elsarticle}

\usepackage{tikz}
\usepackage{pgflibraryarrows}
\usepackage{pgflibrarysnakes}

\usetikzlibrary{trees}

\usepackage[latin1]{inputenc}
\usepackage{tikz}
\usetikzlibrary{trees}
\usetikzlibrary{decorations.pathmorphing}
\usetikzlibrary{decorations.markings}

\usepackage{amsmath}
\usepackage{amssymb}
\usepackage{amsfonts}
\usepackage{theorem}
\usepackage{float}
\usepackage{enumerate}
\usepackage{pgfplots}
\usepackage[plainpages=false,pdfpagelabels=true,colorlinks=true,citecolor=blue,hypertexnames=false]{hyperref}
\usepackage[linesnumbered,ruled,commentsnumbered]{algorithm2e}
\newtheorem{theorem}{Theorem}

\newtheorem{Proposition}[theorem]{Proposition}

\newtheorem{Lemma}[theorem]{Lemma}

\newtheorem{remark}{Remark}
\newtheorem{define}[theorem]{Definition}
\newtheorem{example}[theorem]{Example}

\newenvironment{Proof}{\noindent {\em Proof:\ }}{$\square$}
\newcommand{\SPC}{\hspace*{15pt}}

\def\F{{\mathbb{F}}}

\def\R{{\bf{R}}}
\def\d{{\bf{d}}}

\def\H{{\bf{H}}}

\def\PS{{\mathcal{S}}}

\def\PS{{\bf{P}}}
\def\MS{{\bf{R}}}
\def\QS{{\bf{Q}}}

\def\zero{\hbox{\rm{Zero}}}

\def\class{\hbox{\rm{cls}}}
\def\cls{\hbox{\rm{cls}}}
\def\tdeg{\hbox{\rm{tdeg}}}

\def\lvar{\hbox{\rm{lvar}}}
\def\lm{\hbox{\rm{lm}}}
\def\initial{\hbox{\rm{init}}}

\def\rseq{\hbox{\rm{rseq}}}
\def\arem{\hbox{\rm{arem}}}

\def\A{{{\mathcal{A}}}}

\def\AS{{{\mathcal{A}}}}

\def\T{{{\mathcal{T}}}}

\def\RS{{\bf{R}}}

\def\MS{{\bf{M}}}

\begin{document}

\begin{frontmatter}

\title{On the Efficiency of Solving Boolean Polynomial Systems with the Characteristic Set Method\tnoteref{sponsor}}

\tnotetext[sponsor]{This work was in part supported by the National Natural Science Foundation of China under Grant No. 61977060 and No. 61877058.}

\author{Zhenyu Huang}
\ead{huangzhenyu@iie.ac.cn}

\author{Yao Sun}
\ead{sunyao@iie.ac.cn}

\author{Dongdai Lin}
\ead{ddlin@iie.ac.cn}

\address{SKLOIS, Institute of Information Engineering, Chinese Academy of Sciences, Beijing 100093,  China}


\begin{abstract}
An improved characteristic set algorithm  for solving Boolean polynomial systems is proposed.
This algorithm is based on the idea of converting all the polynomials into monic ones by zero decomposition, and using additions to obtain pseudo-remainders.
Three important techniques are applied in the algorithm. The first one is eliminating variables by new generated linear polynomials. The second one is optimizing the strategy of choosing polynomial for zero decomposition. The third one is to compute  add-remainders to eliminate the leading variable of new generated monic polynomials. By analyzing the depth of the zero decomposition tree, we present some complexity bounds of this algorithm, which are lower than the complexity bounds of previous characteristic set algorithms. Extensive experimental results show that this new algorithm is more efficient than previous characteristic set algorithms for solving Boolean polynomial systems.
%


\end{abstract}


\end{frontmatter}

\section{Introduction}\label{sec_introduction}
Solving Boolean polynomial systems, which is solving polynomial systems in the ring $\R_2=\F_2[x_1,x_2,\ldots,x_n]/\langle x_1^2+x_1,x_2^2+x_2,\ldots,x_n^2+x_n\rangle$,  plays a fundamental role in many important fields such as coding theory, cryptology, and analysis of computer hardware. To find efficient algorithms for solving such systems is a central issue both in mathematics and in computer science.
In the past 20 years, efficient algorithms for solving Boolean polynomial systems have been developed, such as  the Gr\"obner basis algorithm\citep{f4,f5,gb_jssc}, the XL algorithm(equivalent to a sub-optimal
version of F4)\citep{xl1}, the algorithms based on SAT solvers or MILP solvers\citep{satconversion,cryptominisat, milp1}, the fast exhaustive search algorithm\citep{fast_search}, and the hybrid algorithms by combing exhaustive search and Macaulay matrix computation \citep{hybrid_gb,joux_hybrid}. Some of these algorithms had good performance on solving some polynomial
systems generated from cryptanalysis problem\citep{hfe,bivium_faugere, bivium_sat, bivium_gb}.
The asymptotic complexity of solving Boolean polynomial systems, especially the complexity of solving Boolean quadratic polynomial systems,  was well studied.
The complexity of Gr\"obner basis algorithms for solving Boolean polynomial systems was investigated in \citep{bardet1,bardet2} by estimating the degree of regularity of the overdetermined polynomial system $\{\PS, \H\}$, where $\PS=\{f_1,f_2,\ldots,f_m\}$ is the input Boolean polynomial system and $\H = \langle x_1^2+x_1,x_2^2+x_2,\ldots,x_n^2+x_n\rangle$ is the field equations. They presented several complexity bounds of the Gr\"obner basis algorithm for semi-regular quadratic systems when $m/n$ tends to different values. Their results showed that for semi-regular quadratic systems with $n$ variables and $m$ polynomials,  when $m\sim Nn (N>1/4)$, the complexity of the Gr\"obner basis computation is single exponential, when $n<<m<<n^2$, the complexity is sub-exponential, and when $m\sim Nn^2$, with $N$ a constant,  the complexity is polynomial.
Furthermore, in \citep{bardet_comp}, Bardet et al. presented a new algorithm to solving quadratic Boolean polynomial system based on the idea of combining exhaustive search and sparse linear algebra.
Under precise algebraic assumptions which are satisfying with probability very close to 1, the deterministic variant of their algorithm has complexity bounded by $O(2^{0.841n})$ when m = n, while a probabilistic variant of their algorithm (Las Vegas type) has expected complexity $O(2^{0.792n})$, where $m$ is the number of polynomials and $n$ is the number of variables.
Moreover, for input systems without side conditions, the best complexity of solving quadratic Boolean polynomial systems is reached by a fast exhaustive search algorithm in $4\log_2n2^n$ bit operations \citep{fast_search}.

The {\bf characteristic set (CS)} method is a tool for studying
polynomial, algebraic differential, and algebraic difference
equation systems \citep{la1,boul,rody1,chou1,cade10,lift,gluo,hub,kalk1,kapur1,la0,lindd1,moller,wang,wu-basic}.
The idea of the method is reducing equation systems in general form
to equation systems in the form of triangular sets.
Then, the zero set of an equation system can be
decomposed into the union of the zero sets of triangular sets.
With this method, solving an equation system can be reduced to
solving univariate equations in cascaded form.
%
The CS method can also be used to compute the dimension, the degree,
and the order for an equation system, to solve the radical ideal
membership problem, and to prove theorems from elementary and
differential geometries.
In most existing work on CS methods, the common zeros of the equations are taken in an algebraically closed field which is infinite. In \citep{gaof2,cs_fq}, the CS method is extended to solving the polynomial systems in finite fields.  Based on the CS algorithm proposed in \citep{cs_fq}, an algorithm for solving Boolean polynomial systems with noise was proposed and well studied \citep{max_posso_huang, max_posso_tcs}. In \citep{PBCS}, the parallel version of the  CS algorithm proposed in \citep{cs_fq} was presented and  efficiently implemented under the high-performance computing environment.
Research about the complexity of the CS method is rare, and most of the existing results are about the complexity of computing one triangular set \citep{gallo,  szanto}.
The unique result about the complexity of the whole zero decomposition process is that for an input Boolean polynomial system with $n$ variables and $m$ polynomials,  the bit-size complexity of the top-down characteristic set ({\bf TDCS$_2$}) algorithm is bounded by $O(2^{log_2(m)n})$ \citep{cs_fq}.
Since for most problems, $m$ is big hence this bound is much worse than the complexity bounds of the methods introduced in last paragraph.

For the CS method, the bottleneck problem that limits its practical efficiency and causes its high asymptotic complexity is the intermediate expression swell, which is  caused from computing pseudo-remainders by polynomial multiplications. This problem is effectively avoided in the CS algorithm {\bf MFCS} (Multiplication-free CS), which is an algorithm for solving Boolean polynomial systems, proposed in \citep{cs_fq}. Hence the practical efficiency of {\bf MFCS} is much better than {\bf TDCS$_2$}. The mainly idea of {\bf MFCS} is to use the properties of $\F_2$ to convert the initials of all Boolean polynomials into constant $1$ by zero decomposition, then compute pseudo-remainders by addition.
For this algorithm, the principle factor which effects its efficiency becomes the number of branches, that is the number of polynomial sets generated in the zero decomposition process.
However,  in \citep{cs_fq}, the authors only introduced the basic algorithm {\bf MFCS}. Some important techniques, which were used in the implementation of {\bf MFCS} and can greatly reduce the number of branches hence improve the efficiency of the algorithm, were not presented. Moreover, the complexity of {\bf MFCS} is unknown, thus how these techniques influence the complexity of the algorithm is also unknown.
Therefore, the motivation of this paper is to estimate the complexity of the CS algorithms  based on the multiplication-free idea, analyze the variation of the complexity after applying these techniques, improve these techniques,  and finally propose an algorithm which has better theoretical and practical complexity than the previous CS algorithms.

The main contributions of this paper are as follows.
\begin{enumerate}[1)]
\item We propose an algorithm {\bf BCS} by modifying {\bf MFCS} with three major techniques. The first one is adding a simplification process into algorithm, in which we use the linear polynomials
generated in the zero decomposition process to eliminate variables. The second one is using an alternative \texttt{choose} function to determine the order of choosing polynomial for zero decomposition.  The third one is that in the zero decomposition process, instead of adding a polynomial $I+1$ into the current branch and $I$ into the new generated branch, we add the add-remainder of  $I+1$ w.r.t. a monic triangular set $\MS$ into the current branch and the add-remainder of $I$ w.r.t. $\MS$ into the new generated branch, where an add-remainder is the output of a new kind of elimination operation.

\item We define a binary tree called the zero decomposition tree, and  convert the problem of estimating the complexity of {\bf BCS} into two problems. The first one is estimating the complexity of solving one branch of the tree, and the second one is estimating the depth of the tree.
For the first problem, we prove that for {\bf BCS} the complexity of solving one branch is bounded by $O(dm(m+b)\log(n)n^{d+3})$, where $d$ is the degree of the system, $m$ is the number of polynomials, $n$ is the number of variables, and $b$ is the depth of the zero decomposition tree.
For the second problem,  we show that for any \texttt{choose}, $b$ is bounded by $m\sum_{j=1}^{d-1}{n \choose j}$. Moreover, we introduce a vector index based on the lowest degree of the non-monic polynomials and monic polynomials, then by analyzing the variation of the sum of the entries in this vector index after zero decomposition and other operations, we show that $b$ is bounded by $(d-1)n$, if \texttt{choose} always choose the polynomial with lowest degree, or the polynomial whose initial has lowest degree.
Then for any input system, the complexity of {\bf BCS} is bounded by $O(dm(m+d-1)\log(n)n^{d+3}2^{(d-1)n})$. Especially, for quadratic polynomial systems, the complexity of {\bf BCS} is bounded by $O(n(m+n)2^n)$. 

\item We implemented {\bf BCS} with C language and tested its efficiency by solving several groups of Boolean polynomial systems generated from cryptanalysis and reasoning problems, and some random generated Boolean polynomial systems.
We compared the timings of {\bf BCS} and other algorithms, such as the {\bf MFCS}  algorithm used in the experiments of \citep{cs_fq}, the  Boolean Gr\"obner basis routine in Magma V2.20, the Boolean Gr\"obner basis routine implemented by the Polybori library in SAGE V8.7, and  a SAT-solver Cryptominisat  V5.6.8.  From the experimental results, we can observe that {\bf BCS} is the most efficient algorithm for solving these problems generated from cryptanalysis and reasoning, and is compare with other algorithms for solving  these random generated problems.
\end{enumerate}

The rest of this paper is organized as follows. In Section 2, we introduce the problem of solving Boolean polynomial systems.  In Section 3, the {\bf BCS} algorithm is proposed. In Section 4, we prove the correctness of {\bf BCS}. In Section 5, we present some complexity bounds of {\bf BCS}. In Section 6, we present the experimental results. In Section 7, conclusions are give.


\section{The problem of Solving Boolean Polynomial Systems}

Let $\F_2$ be the finite field of two elements $\{0,1\}$, and $\{x_1,x_2,\ldots,x_n\}$ be a variable set. Unless otherwise stated,  these variables are ordered as $x_1< x_2 < \cdots < x_n$. Consider the Boolean polynomial ring $\R_2= \F_2[x_1,x_2,\ldots,x_n]/\langle x_1^2+x_1, x_2^2+x_2, \ldots, x_n^2+x_n\rangle$. An element in $\R_2$ is called a Boolean polynomial.
In this paper, we consider the following problem.

\vspace{5pt}

{\bf The Boolean Polynomial Systems Solving  (Boolean PoSSo) Problem}

{\bf Input:} A Boolean polynomial set $\PS=\{f_1,f_2,\ldots,f_m\}\subseteq\R_2$.

{\bf Output:} Solutions $(x_1, x_2,\ldots, x_n)\in \F_2^n$ s.t.  $f_i(x_1,x_2,\ldots,x_n)=0$ for $i=1,2,\ldots,m$.

\vspace{5pt}

Note that, the solutions of Boolean PoSSo problems are all restrained in the filed $\F^n_2$. Example \ref{ex_toy} is a simple example for this problem, and we will use it to illustrate the procedures of our algorithm in the following parts of this paper.

\begin{example}\label{ex_toy}
Given a Boolean polynomial system $\PS=\{f_1,f_2,f_3\}=\{ (x_4+x_1x_2x_3+x_1x_3)x_5+(x_2+1)x_3x_4+x_1x_2x_3+x_2+1, (x_4+x_1x_2x_3+x_2+1)x_5+(x_1x_2x_3+x_1)x_4+x_2x_3+x_1, x_1x_3x_4x_5+(x_1x_3+x_2)x_4+x_3+x_2+x_1\}$.
The solutions of $\PS$ for the Boolean PoSSo problem are (1, 0, 1, 0, 1),  (1, 0, 1, 1, 1),  (1, 1, 1, 1, 1), (0, 1, 0, 1, 0), (0, 0, 0, 1, 1).
\end{example}

In the following paragraphs of this paper, unless otherwise stated, for a polynomial system, we mean a Boolean polynomial system. Moreover, we use $n$ to denote the number of variables and $m$ to denote the number of polynomials.

Now we show how to use the characteristic algorithms to solve the Boolean PoSSo problem. First, we introduce some basic notions and notations about the characteristic set method over $\F_2$. For more details, the reader is referred to \citep{cs_fq}.

For a Boolean polynomial $P\in \R_2$,
the {\em class} of $P$, denoted as $\cls(P)$,  is the largest index $c$ such that $x_c$ occurs in $P$.
If $P$ is a constant, we set $\cls(P)$ to be $0$.
If $\class(P)=c>0$, we call $x_c$ the {\em
leading variable} of $P$, denoted as $\lvar(P)$.
The leading coefficient of $P$ as a univariate polynomial  in
$\lvar(P)$ is called the {\em initial} of $P$, and is denoted as
$\initial(P)$. Then, $P$ can be written as $Ix_c+U$, where $I=\initial(P), c=\cls(P)$, and $U$ is a polynomial without variables $x_c$.

Given a polynomial system $\PS$, we use $\zero(\PS)$ to denote the solutions of this system, and we call $\zero(\PS)$ the zero set of $\PS$.

A sequence of nonzero polynomials
\begin{equation}\label{eq-a}\nonumber
\AS:~~ A_1, A_2,\ldots, A_r
\end{equation}
is a {\em triangular set} if either $r=1$ and $A_1=1$, or
$0<\cls(A_1) < \cdots < \cls(A_r)$. A Boolean polynomial $P$ is called {\em monic}, if $\initial(P)=1$.
Moreover, if the elements of a triangular set are all monic, we call it a monic triangular set.
Given a monic triangular set $\AS: A_1, A_2,\ldots, A_r$, we can easily obtain its zero sets. Specifically speaking,  for each evaluation of the variables in $\{x_1,x_2,\ldots,x_n\}\setminus \{\lvar(A_1),\lvar(A_2),\ldots, \lvar(A_r)\}$, we can obtain one solution in $\zero(\A)$ by recursive substitution.  Hence, we have $|\zero(\AS)|=2^{n-r}$, and $n-r$ is called the {\em dimension} of $\A$.

The characteristic set algorithms proposed in \citep{cs_fq} is to solve the following zero decomposition problem.

\vspace{5pt}

{\bf The Boolean Zero Decomposition Problem}

{\bf Input:} A Boolean polynomial set $\PS=\{f_1,f_2,\ldots,f_m\}\subseteq\R_2$.

{\bf Output:} Monic triangular sets $\A_1, \A_2, \ldots, \A_r$, such that $\zero(\PS)=\bigcup_i\zero(\A_i)$,  and $\zero(\A_i)\cap \zero(\A_j)=\emptyset$ for any $i\neq j$.

\vspace{5pt}

Obviously, if one can obtain such monic triangular sets $\A_i$,  the solutions of $\PS$ can be easily achieved, hence the corresponding Boolean PoSSo problem is solved. Moreover, $|\zero(\PS)|=\sum_i2^{n-r_i}$, where $r_i$ is the number of polynomials in $\A_i$.

\begin{example}
For the polynomial system $\PS$ in Example \ref{ex_toy}. By the characteristic set method, we can obtain five monic triangular sets $\A_1=\{x_1+1, x_2, x_3+x_2+1, x_4, x_5+x_2+1\}, \A_2=\{x_1+1, x_2, x_3+1, x_4+1, x_5+1\}, \A_3=\{x_1+1, x_2+1, x_3+x_2, x_4+1, x_5+1\}, \A_4=\{x_1, x_2+1, x_3, x_4+1, x_5\}, \A_5=\{x_1, x_2, x_3, x_4+1, x_5+1\}$, which are the outputs of the Boolean zero decomposition problem\footnote{By different algorithms, the output triangular sets may be different.}.

\end{example}

\section{An  improved Characteristic Set algorithm}\label{sec_algorithm}

In this section, we will introduce the improved characteristic set algorithm {\bf BCS}.
First, we define a new kind of elimination operation. In the following of this paper, we use $\tdeg(P)$ to denote the total degree of a polynomial, and $\lm(P)$ to denote the leading monomial of $P$ w.r.t. a monomial order \citep{cox_book}.

\begin{define}
Suppose  $P$ is a  polynomial and $\MS$ is a monic triangular set.
Let $P'$ and $\MS'$ be the outputs of the following process.
\vspace{5pt}

Step 1. Set $P'$ to be $P$ and $\MS'$ to be $\MS$.


Step 2. While $P'$ is monic and nonlinear, and $\exists Q\in \MS'$ s.t. $\cls(P')=\cls(Q)$, do:

\hspace{10pt} Step 2.1. If $\tdeg(P')<\tdeg(Q)$, then replace $Q$ with $P'$ in $\MS'$;

\hspace{10pt} Step 2.2. Set $P'$ to be $P'+Q$;

Step 3. Output $P'$ and $\MS'$.
\vspace{5pt}

We call $P'$ the {\bf add-remainder} of $P$ w.r.t $\MS$, denoted by $\arem(P, \MS)$. $\MS'$ is called the {\bf reducer sequence}  of $P$ w.r.t. $\MS$, denoted by $\rseq(P, \MS)$.

\end{define}

Obviously,  we have $\zero(P, \MS) = \zero(\arem(P, \MS), \rseq(P, \MS))$.
Now we show the precise process of {\bf BCS}

\

{
\IncMargin{1em}
\begin{algorithm}[H] \label{alg-BCS} \caption{\bf BCS}
\SetAlgoVlined
\SetKwFunction{basic}{Simplify}\SetKwFunction{triset}{Triset}
\SetKwInOut{input}{input}\SetKwInOut{output}{output}
\input{A polynomial system $\PS=\{f_1,f_2,\ldots,f_m\}$.}
\output{Monic triangular sets $\{\A_1,\A_2,\ldots,\A_t\}$ such that
 $\zero(\PS)=\cup_{i=1}^t\zero(\A_i)$ and $\zero(\A_i)\cap \zero(\A_j)=\emptyset$
}
  \noindent
 $\PS^*\leftarrow \{\PS\}, \A^*\leftarrow \emptyset$\;
 \While(\tcc*[f]{$\PS^*$ is a group of polynomial sets}){$\PS^*\ne \emptyset$}
 { Select and remove a polynomial set $\QS$ from $\PS^*$\;
  Let $\A$ and $\QS^*$ be the output of {\bf Triset}({$\QS$})\;
  \lIf{$\A\ne \emptyset$}{$\A^*\leftarrow \A^*\cup\{\A\}$}
  $\PS^*\leftarrow \PS^*\cup \QS^*$\;
 }
 \Return $\A^*$.
 \end{algorithm}
\DecMargin{1em}

\newpage

\DecMargin{1em}
\IncMargin{1em}
\begin{algorithm}[H]
\NoCaptionOfAlgo\SetAlgoVlined
\caption {Function: \texttt{Simplify}}\label{func_simplify}
\SetKwInOut{input}{input}\SetKwInOut{output}{output}
\input{ A polynomial set $\PS$ and a monic polynomial set $\MS$.}
\output{ A linear triangular set $\A'$, and a nonlinear polynomial set $\PS'$ and a monic polynomial set $\MS'$,  such that $\zero(\PS)\cup \zero(\MS)=\zero(\A') \cup \zero(\PS')\cup \zero(\MS').$}
\BlankLine
$\A'\leftarrow \emptyset, \PS'\leftarrow \PS, \MS' \leftarrow \MS$ \;
\lIf{$1 \in \PS'$}{\Return $\emptyset$, $\emptyset$ and $\emptyset$}
\While(\tcc*[f]{$\cls(P)=c$}){$\PS'$ has a linear polynomial $P=x_c+L$}
{
$\PS'=\PS' \setminus\{P\}$\;
Substitute $x_c$ with $L$ for the other polynomials in $\PS'$\;
Suppose $\MS'=\{M_1,M_2,\ldots, M_k\}$\;
\For{$i\leftarrow 1$ to $k$}
{
  Substitute $x_c$ with $L$ in $M_i$, and obtain $M'_i$\;
  $\MS'=\MS'\setminus \{M_i\}$\;
  \lIf{$\cls(M'_i)=\cls(M_i)$ and $M'_i$ is not linear}{ $\MS'\leftarrow \MS' \cup \{M'_i\}$}
  \lElse{$\PS' = \PS' \cup \{M'_i\}$}
}
$\A'\leftarrow \A' \cup \{x_c+L\}$\;
\lIf{$1 \in \PS'$}{\Return $\emptyset$, $\emptyset$, and $\emptyset$}
}
 \Return $\A'$, $\PS'$ and $\MS'$.
\end{algorithm}

\

\DecMargin{1em}
\IncMargin{1em}
\begin{algorithm}[H]
\NoCaptionOfAlgo\SetAlgoVlined
\caption {Function: \texttt{AddReduce}}\label{func_addreduce}
\SetKwInOut{input}{input}\SetKwInOut{output}{output}
\input{ A monic polynomial set $\PS$.}
\output{ A polynomial set $\PS'$, and a monic triangular set $\MS'$,
such that $\zero(\PS) = \zero(\PS') \cup \zero(\MS')$.}
\BlankLine
$\PS'\leftarrow \emptyset$\;
\Repeat{$\RS = \emptyset$}
{
 Sort the elements of $\PS$ by classes and obtain polynomial sets $\QS_1,\QS_2,\ldots,\QS_t$
\tcc*{The class of elements in $\QS_i$ is $c_i$}
 $\MS' \leftarrow \emptyset$, $\RS \leftarrow \emptyset$\;
\For{$i\leftarrow 1$ to $t$}
{
Let $Q$ be a polynomial in $\QS_i$, such that $\lm(Q)$ is smallest w.r.t a graded order
\;
$\QS_i\leftarrow \QS_i\setminus \{Q\}$, $\MS'=\MS'\cup \{Q\}$\;
\While{$\QS_i\ne \emptyset$}
{ Choose an element $Q_j\in \QS_i$, $\QS_i\leftarrow \QS_i\setminus \{Q_j\}$\;
 $Q_j\leftarrow Q_j+Q$\;
 \lIf{$ Q_j=1$}{\Return $\emptyset$ and $\emptyset$}
 \lIf{$Q_j \ne 0$, and $Q_j$ is linear or not monic}{$\PS' \leftarrow \PS'\cup \{Q_j\}$}
 \lIf{$Q_j$ is monic}{ $\RS\leftarrow \RS\cup \{Q_j\}$}
  }
}
\lIf{$\RS\ne \emptyset$}{$\PS\leftarrow \MS'\cup \RS$}
}

\Return $\PS'$ and $\MS'$.

\end{algorithm}

}

\setcounter{algocf}{1}

\IncMargin{1em}
\begin{algorithm}\label{alg-triset} \caption{\bf TriSet}
\SetAlgoVlined
\SetKwFunction{simplify}{Simplify}\SetKwFunction{addreduce}{AddReduce}
\SetKwFunction{fchoose}{Choose}
\SetKwInOut{input}{input}\SetKwInOut{output}{output}
\input{A polynomial system $\PS=\{f_1,f_2,\ldots,f_m\}$.}
\output{A monic triangular set $\A$ and a group of polynomial sets $\QS^*$ such that
$\zero(\PS)=\zero(\A)\cup_{\QS\in \QS^*} \zero(\QS)$, $\zero(\QS_i)\cap \zero(\A)=\emptyset$,
$\zero(\QS_i)\cap \zero(\QS_j)=\emptyset$ for any $\QS_i,\QS_j\in \QS^*$ with $i\ne j$.
}
  \noindent
 $\QS^*\leftarrow \emptyset$, $\A\leftarrow \emptyset$, $\MS \leftarrow \emptyset$ \;
 \While{$\PS\ne \emptyset$}
{

\Repeat{$\PS$ doesn't contain linear polynomials}
 {
   Let $\A'$, $\PS'$, $\MS'$ be the output of  $\simplify(\PS, \MS)$\;
  \lIf{$\A',\PS', \MS'=\emptyset$}{\Return $\emptyset$ and $\QS^*$}
  \lElse{$\A\leftarrow \A\cup\A'$, $\PS\leftarrow \PS'$, $\MS \leftarrow \MS'$}

   Let $\MS'$ be the set of all monic polynomials in $\PS$\;
  $\PS\leftarrow \PS\setminus \MS'$, $\MS \leftarrow \MS \cup \MS'$\;
\If{$\MS \ne \emptyset$}
 {
  Let $\MS'$ and $\PS'$ be the output of $\addreduce{$\MS$}$\;
  \lIf{$\MS'=\emptyset$}{\Return $\emptyset$ and $\QS^*$}
  \lElse{$\PS\leftarrow \PS\cup \PS'$, and $\MS \leftarrow \MS'$}
 }
  }
\If(\tcc*[f]{The zero decomposition process}){$\PS \ne \emptyset$}
{
$P\leftarrow$ \fchoose{$\PS$},  and suppose $\cls(P)=k, P=Ix_k+U$\tcc*{\fchoose is a function that chooses an element from a polynomial set}
\eIf{$\arem(I, \MS)$ is not a constant}
{
$I \leftarrow \arem(I, \MS)$, $\MS \leftarrow \rseq(I, \MS)$\;
$\PS_1\leftarrow (\PS\setminus \{P\})\cup\A \cup \MS \cup\{I,U\}$, and
 $\QS^*\leftarrow \QS^*\cup\{\PS_1\}$\;
 $\PS\leftarrow (\PS\setminus \{P\})\cup\{x_c+U\}$\;
 \lIf{$I$ is monic and nonlinear}{$\MS\leftarrow \MS\cup\{I+1\}$}
 \lElse{$\PS\leftarrow \PS\cup\{I+1\}$}
}
{
\lIf{$\arem(I, \MS) = 1$}
{$\PS\leftarrow \PS\setminus \{P\} \cup \{x_c+U\}$
}
\lElse{$\PS\leftarrow \PS\setminus \{P\} \cup \{U\}$}
\While{$\arem(I, \MS)$ is a constant }
{
Suppose $I=x_{c_0}+x_{c_1}+\cdots+x_{c_k}+I'x_p+U'$, where $I' \ne 1$ and $c_0>c_1>\cdots>c_k>p>\cls(U')$\;
$I\leftarrow I'$\;
}
$I \leftarrow \arem(I, \MS)$, $\MS \leftarrow \rseq(I, \MS)$\;
$\PS_1\leftarrow \PS\cup\A \cup \MS \cup\{I\}$, and
 $\QS^*\leftarrow \QS^*\cup\{\PS_1\}$\;
 \lIf{$I$ is monic and nonlinear}{$\MS\leftarrow \MS\cup\{I+1\}$}
 \lElse{$\PS\leftarrow \PS\cup\{I+1\}$}
}
}
}

$\A \leftarrow \A\cup \MS$,  and \Return $\A$ and $\QS^*$.
\end{algorithm}

\newpage

In {\bf BCS}, {\bf Triset} is the sub-algorithm that solves the current polynomial system and generates some new polynomial systems, and it is the major part of {\bf BCS}. Here we explain several main processes of {\bf Triset}.
\begin{itemize}
\item At Step 3-13, we are trying to find linear polynomials and monic polynomials in the current system. If there is a linear polynomial $x_c+L$, we use $L$ to substitute $x_c$ in other polynomials, and move this $x_c+L$ into the monic triangular set $\A$. For the monic polynomials, we execute \addreduce to eliminate the leading variables of those polynomials which have the same classes by addition.

\item At Step 16-21, we convert the chosen polynomial into monic polynomial by zero decomposition. It is based on the fact that $\zero(Ix_c+U)=\zero(x_c+U,I+1)\cup \zero(I,U)$. Then we compute $\arem(I, \MS)$ to simplify $I$. $\arem(I, \MS)$ is added into the new generated polynomial set. If $\arem(I, \MS)$ is monic and nonlinear, we add $\arem(I, \MS)+1$ into $\MS$. Otherwise, we add $\arem(I, \MS)+1$ into $\PS$.

\item At Step 22-31, $\arem(I, \MS)$ is equal to a constant $c \in \F_2$\footnote{The probability of this case is extremely low from the observation in  our experiments}. Obviously, we have $I+M_1+M_2+\cdots+M_k=c$ for some $M_1,\ldots, M_k\in \MS$, then $\zero(Ix_c+U, \MS)=\zero(c x_c+U, \MS)$. Hence, we replace $Ix_c+U$ with $c x_c+U$ in $\PS$.
Note that $\tdeg(I)<\tdeg(Ix_c+U)$, and we want to well use this polynomial with lower degree in the following process. However, when $\arem(I, \MS)$ is constant, $I$ is equivalent to a constant and we cannot achieve a polynomial with lower degree. Hence, by step 25-27,  we generate a  lower degree polynomial $I'$ from $I$, then do the zero decomposition based on the cases of $\arem(I', \MS)=0$ or $1$.

\end{itemize}

As mentioned before, {\bf BCS} is originated from the {\bf MFCS} algorithm proposed in \citep{cs_fq}.
The similarity  of {\bf BCS} and {\bf MFCS} is the idea of using addition to eliminate variables.
These two algorithms have four major differences:
\begin{enumerate}[1)]
\item In {\bf MFCS}, \fchoose always chooses the polynomials with highest class. In {\bf BCS}, \fchoose can be any form.
\item In {\bf MFCS}, \addreduce is executed when the polynomials with the highest class are all monic. In
{\bf BCS}, \addreduce is executed when we have new generated monic polynomials.
\item In {\bf BCS}, we add a new function \simplify to deal with the linear polynomials generated in the solving process.
\item In {\bf BCS}, we use $\arem(I, \MS)$ instead of $I$ in the zero decomposition processes.
\end{enumerate}

As we mentioned in Section \ref{sec_introduction},  in the experiments of \citep{cs_fq}, some techniques were already added in the implementation of {\bf MFCS}, here we denote this modified algorithm by {\bf MFCS}$_1$.
In {\bf MFCS}$_1$, $\simplify$ is used and \fchoose always chooses the polynomial whose initial is shortest, that is choosing the polynomial whose initial has the smallest number of monomials.
Hence,  {\bf BCS} and {\bf MFCS}$_1$ have the above differences 1), 2), 4).

From the complexity analysis in Section \ref{sec_complexity}, we will see that these differences are important to reducing the complexity of characteristic set algorithms. Moreover, experimental results in Section \ref{sec_experiment} shows that by this modifications {\bf BCS} is more efficient that {\bf MFCS}$_1$ in practical computations.


%

\begin{example}\label{ex_bcs}
Let the polynomial system $\PS$ in Example \ref{ex_toy} be the input of {\bf BCS}. We suppose that by \fchoose we always choose the polynomial with lowest degree. We show the procedure of {\bf BCS}(\PS) step by step.
\begin{enumerate}
\item First, we let $\PS$ be the input of {\bf Triset}, and choose $f_1$ to do zero decomposition. The initial of $f_1$ is $I_1=x_4+x_1x_2x_3+x_1x_3$. Then, after zero decomposition, $\PS$ is updated to be $\PS^1=\{f_4,  f_5,  f_2, f_3\}$, where $f_4=I_1+1, f_5=x_5+(x_2+1)x_3x_4+x_1x_2x_3+x_2+1$. We generate a new polynomial set $\PS_1=\{I_1, g_1, f_2, f_3\}$, where $g_1=(x_2+1)x_3x_4+x_1x_2x_3+x_2+1$. Note that $f_4$ and $f_5$ is monic, hence $\MS=\{f_4, f_5\}$. Now we choose $f_2$ to do zero decomposition. The initial of $f_2$ is $I_2=x_4+x_1x_2x_3+x_2+1$, which is a monic polynomial. Thus, $\arem(I_2, \MS)= I_2+f_4=x_1x_3+x_2$, and $\rseq(I_2, \MS)=\MS$. We set $I'_2$ to be $x_1x_3+x_2$. Hence, $\PS^1$ is updated to be $\PS^2=\{f_6, f_7, f_4, f_5, f_3\}$, where $f_6= I'_2+1$ and $f_7=x_5+(x_1x_2x_3+x_1)x_4+x_2x_3+x_1$. We generate a new polynomial set $\PS_2=\{I'_2, g_2, f_4, f_5, f_3\}$, where $g_2=(x_1x_2x_3+x_1)x_4+x_2x_3+x_1$. In $\PS^2$, there are three monic polynomials $f_7, f_4, f_5$. Moreover, $f_7$ and $f_5$ have the same class. So after \addreduce, $f_7$ is reduced to $f'_7=f_7+f_5=(x_1x_2x_3+x_2x_3+x_3+x_1)x_4+(x_1x_2+x_2)x_3+x_2+x_1+1$. Then, we choose $f_6=x_1x_3+x_2+1$ to do zero decomposition. Its initial is $I_3=x_1$, hence $\PS^2$ is updated to be $\PS^3= \{x_1+1, x_3+x_2+1, f'_7, f_4, f_5, f_3\}$, and we generate a new polynomial set $\PS_3=\{x_1, x_2+1, f'_7, f_4, f_5, f_3\}$. Since $x_1+1$ and $x_3+x_2+1$ are linear, we can execute \simplify, and after that, we obtain a linear triangular set $\A_1=\{x_1+1, x_2, x_3+x_2+1, x_4, x_5+x_2+1\}$ and $\PS^3$ becomes a empty set. Hence, {\bf Triset} outputs $\A_1$ and $\{\PS_1, \PS_2, \PS_3\}$.

\item Let $\PS_1=\{I_1, g_1, f_2, f_3\}$ be the input of {\bf Triset}. We choose $g_1$ to do zero decomposition. The initial of $g_1$ is $I_4=(x_2+1)x_3$. Then, $\PS_1$ is updated to be $\PS^1_1=\{p_1, p_2, I_1, f_2, f_3\}$, where $p_1=I_4+1$ and $p_2=x_4+x_1x_2x_3+x_2+1$. We generate a new polynomial set $\PS_4=\{I_4, p_3, I_1, f_2, f_3\}$, where $p_3 = x_1x_2x_3+x_2+1$. In $\PS_1^1$, we find that $I_1$ and $p_2$ are monic and have the same class. Thus by \addreduce, we obtain a new polynomial $p_4 = p_2+I_1=x_1x_3+x_2+1$, and replace $I_1$ with $p_4$ in $\PS_1^1$. Now $\PS_1^1=\{p_1,p_4, f_2, f_3\}$ and $\MS=\{p_2\}$. We choose $p_1$ to continue the zero decomposition. The initial of $p_1$ is $I_5= x_2+1$, hence $\PS_1^1$ is updated to $\PS_1^2=\{x_2, x_3+1, p_4, f_2, f_3, p_2\}$, and we generate a new polynomial set $\PS_5=\{x_2+1, 1, p_4, f_2, f_3, p_2\}$. Now, we execute \simplify for $\PS_1^2$. After that, the linear polynomial set $\A_2$ is $\{x_1+1, x_2, x_3+1,  x_4+1, x_5+1\}$ and $\PS_1^2=\emptyset$. {\bf Triset} outputs $\A_2$ and $\{\PS_4, \PS_5\}$.

\item  Let $\PS_2=\{I'_2, g_2, f_4, f_5, f_3\}$ be the input of {\bf Triset}. We choose $I'_2$ whose initial is $I_6=x_1$  to do zero decomposition. Obviously, after zero decomposition, $\PS_2$ is updated to be $\PS_2^1=\{x_1+1, x_3+x_2, g_2, f_4, f_5, f_3\}$, and a new polynomial set $\PS_6= \{x_1, x_2, g_2, f_4, f_5, f_3\}$ is generated. Now we execute \simplify for $\PS_2^1$. Then, we can obtain a linear triangular set $\A_3=\{x_1+1, x_2+1, x_3+x_2, x_4+1, x+5+1\}$. Thus, the output of {\bf Triset} is $\A$ and $\{\PS_6\}$.

\item Let $\PS_3=\{x_1, x_2+1, f'_7, f_4, f_5, f_3\}$ be the input of {\bf Triset}. After \simplify, we have $\A_4=\{x_1, x_2+1, x_3, x_4+1, x_5\}$ and $\PS_3=\emptyset$. Hence, {\bf Triset} outputs $\A_4$ and $\emptyset$.

\item Let $\PS_4=\{I_4, p_3, I_1, f_2, f_3\}$ be the input of {\bf Triset}. We choose $I_4$ to do zero decomposition. Its initial is $I_7=x_2+1$. Then $\PS_4$ is updated to be $\PS_4^1=\{x_2, x_3, p_3, I_1, f_2, f_3\}$, and we generate a new polynomial set $\PS_7=\{x_2+1, p_3, I_1, f_2, f_3\}$. After \simplify, $p_3$ is reduced to constant 1, thus {\bf Triset} outputs $\emptyset$ and $\PS_7$.

\item Let $\PS_5=\{x_2+1, 1, p_4, f_2, f_3, p_2\}$ be the input of {\bf Triset}. Obviously, constant 1 is in $\PS_5$, hence the output are two empty sets.

\item Let $\PS_6= \{x_1, x_2, g_2, f_4, f_5, f_3\}$ be the input of {\bf Triset}. After \simplify, we have $\A_5=\{x_1, x_2, x_3, x_4+1, x_5+1\}$, and {\bf Triset} outputs $\A_5$ and $\emptyset$.

\item Let $\PS_7=\{x_2+1, p_3, I_1, f_2, f_3\}$ be the input of {\bf Triset}. After \simplify, we have $f_3$ is reduced to constant $1$, thus {\bf Triset} outputs two empty sets.

\end{enumerate}

Finally, {\bf BCS} outputs five monic triangular sets $\{\A_1, \A_2, \A_3, \A_4, \A_5\}$.
\end{example}

\section{The correctness of {\bf BCS}}

In this section, we will prove the correctness of {\bf BCS}.
In the following of this paper, when we say the first kind of zero decomposition,  we mean the procedure of  step 16- 21 in  {\bf Triset}, and when we say the second kind of zero decomposition, we mean the procedure of step 22-31 in {\bf Triset}.

\begin{Lemma}\label{lm_triset}
Algorithm {\bf Triset} is correct.
\end{Lemma}
\begin{Proof}
It is easy to check that \addreduce and \simplify is correct.
Let $\PS_0$ be $\PS\cup \A \cup \MS$, which is the polynomial systems we deal with in Loop 2.
Obviously, the elements in $\PS_0$ may be updated after each iteration of Loop $2$.
Since $\zero(P,Q)=\zero(P,P+Q)$ and $\zero(Ix_c+U,x_c+L)=\zero(IL+U,x_c+L)$,  we know that except the zero decomposition operations, other operations will not change the zero set of $\PS_0$.
Now we consider the first kind of zero decomposition, it is based on the fact that
$\zero(\PS_0)=\zero(\PS_0\setminus \{Ix_c+U\}, x_c+U,I+1)\cup \zero(\PS_0\setminus \{Ix_c+U\}, I,U)$ and $\zero(\PS_0\setminus \{Ix_c+U\}, x_c+U,I+1)\cap \zero(\PS_0\setminus \{Ix_c+U\}, I,U)=\emptyset$.
Obviously, if we respectively use $\arem(I, \MS), \rseq(I, \MS)$ instead of $I , M$, the above equations are still valid.
Consider the second kind of zero decomposition. It happens when $\arem(I, \MS)$ is a constant.
Then $\zero(\PS_0)=\zero(\PS_0\setminus\{Ix_c+U\}, x_c+U)$ when $\arem(I, \MS)=1$, and $\zero(\PS_0)=\zero(\PS_0\setminus\{Ix_c+U\}, U)$ when $\arem(I, \MS)=0$. By Step 25-27, a new non-constant polynomial $I$ is generated.  Then $\zero(\PS_0) =\zero(\PS_0, \arem(I, M)+1)\cup \zero(\PS_0, \arem(I, M))$, and $\zero(\PS_0, \arem(I, M)+1)\cap \zero(\PS_0, \arem(I, M))=\emptyset$.
The above equations show that if the algorithm outputs the result, we have $\zero(\A)\cup_{\QS\in \QS*} \zero(\QS^*)=\zero(\PS)$.  This proves the correctness of the zero decomposition equations of the output.

Now we show that $\A$ must be a monic triangular set. There are two cases in which a polynomial $P$ can be added into $\A$.
\begin{enumerate}[1)]
\item $P=x_c+L$ is a linear polynomial. It is monic, and we substitute all the $x_c$ with $L$ in other polynomials, hence other elements in $\A$ will not have $x_c$.

\item $P\in \MS$ is added into $\A$ at step $32$. In this case,  for every class, there is only one element  in $\MS$ and it is monic.  Therefore, the elements being added into $\A$ will have different class.
\end{enumerate}

\noindent It implies that the elements in $\A$ are monic and have different classes, which means $\A$ is a monic triangular set.

Now let's prove the termination of {\bf Triset}. It is sufficient to show that the loop of Step $2$ terminates.
We prove this by induction. If $n=1$, the termination is obvious. Now we assume when $n\le k$, Loop 2 terminates. When $n=k+1$, we prove that if no contradiction occurs, which means we don't obtain constant $1$, we will convert all the polynomials with class $k+1$ into monic ones.
Suppose the \fchoose function doesn't choose the polynomial with class $k+1$,
then we will always deal with the polynomials with $k$ variables.
According to the hypothesis, we will find contradiction or achieve a monic polynomial set with different
classes from these polynomials. If no contradiction occurs, then the \fchoose function have to choose the polynomial with class $k+1$. The above procedure will repeat until all the polynomials with class $k+1$ are converted into monic ones.  Then, after executing \addreduce, $\PS$ will have one polynomials with class $k+1$, and it is monic. After that, we only need to deal with the polynomials with $k$ variables, thus Loop 2 will terminate at last.
\end{Proof}

\begin{theorem}
Algorithm {\bf BCS} is correct.
\end{theorem}
\begin{Proof}
According to Lemma \ref{lm_triset}, it is easy to check that if {\bf BCS} terminates, the output is correct. Therefore, we only need to prove the termination of {\bf BCS}.

For a polynomial set in $\PS^*$, we assign an index $(t_n,r_n,t_{n-1},r_{n-1},\ldots,t_1,r_1)$, where $t_i$ is the number of non-monic polynomials with class $i$ in $\PS^*$, and $r_i$ is the number of polynomials with class $i$ in $\PS^*$. Then we order the indexes by lexicography.
Now we will show that the index of any polynomial set in $\QS^*$ is strictly smaller than the index of $\QS$.
Let $\PS_0=\PS\cup \MS\cup \A$ be the current polynomial set in {\bf Triset}.
It is sufficient to show that in {\bf Triset}, the indexes of $\PS_0$ will not increase after $\PS_0$ being updated,  and the index of a new generated $\PS_1$ is always smaller than that of $\PS_0$ before zero decomposition.

We consider the following four kinds of operations.
\begin{enumerate}[1)]
\item We execute \addreduce. Consider the monic polynomials with highest class $c$. Obviously,  after \addreduce, $t_c$ will not be change, and $r_c$ will decrease. Thus, the index of $\PS_0$ will decrease.

\item We execute \simplify. Suppose we have a linear polynomial $x_c+l$. Note that only polynomials with class not less than $c$ will be changed after substitution. For these polynomials, if some of them are converted into polynomials with lower classes, then the index of $\PS_0$ will decrease. If the classes of them are not changed after substitution, then only $t_k$ with $k>c$ may decrease when some non-monic polynomials are converted into monic ones.  This implies that the index of $\PS_0$ will not increase.

\item We choose a non-monic polynomial $Ix_c+U$, where $\arem(I, \MS)$ is not a constant, to do zero decomposition. Then we replace $Ix_c+U$  by $x_c+U$ in $\PS_0$, add $\arem(I,\MS)+1$, whose class is lower than $c$,  into $\PS_0$, and replace $\MS$ by $\rseq(I, \MS)$ in $\PS_0$. Note that, replacing $\MS$ by $\rseq(I, \MS)$ will not change the index of $\PS_0$. Hence, in this case, $t_c$, the number of non-monic polynomial with class $c$,  decrease by $1$, then the index of $\PS_0$ will decrease. Moreover, in the new generated polynomial set $\PS_1$, $Ix_c+U$ is replaced by $U$ and $\arem(I, \MS)$, hence the index of $\PS_1$ is lower than that of $\PS_0$ before zero decomposition.

\item We choose a non-monic polynomial $Ix_c+U$, where $\arem(I,\MS)$ is a constant,  to do zero-decomposition. $Ix_c+U$ is replace by $x_c+U$ or $U$, and a new polynomial whose class is lower than $c$ is added into $\PS_0$. Thus  $t_{c}$ or $r_{c} $decrease by 1, which means the index of $\PS_0$ will decrease. For the new generated polynomial set $\PS_1$, obviously its index is equal to the index of the updated $\PS_0$, hence is lower than the index of $\PS_0$ before zero decomposition.

\end{enumerate}


It is easy to show that a strictly decreasing sequence
of indexes must be finite. This proves the termination of {\bf BCS}.
\end{Proof}

\section{The complexity of {\bf BCS}}\label{sec_complexity}

In this section, we will estimate the complexity of {\bf BCS}. In order to do this, we introduce the concept of zero decomposition tree.
We can generate {\em the zero decomposition tree} of {\bf BCS} as follows.

\begin{enumerate}[I)]
\item First, let the root node to be the input polynomial system $\PS$, and let a pointer $\mathcal{M}$ point to this node.  The depth of the root node is set to be $0$.

\item In the process of {\bf Triset}, when we generate a new polynomial system by zero decomposition, we generate a new node, and set it  to be this new polynomial system. Then let this node be the right child of the node pointed by $\mathcal{M}$.
After we update the current polynomial system after zero decomposition, we generate a new node, and set it to be the updated polynomial system. Then let this node to be the left child of the node pointed by $\mathcal{M}$.
After this, we let $\mathcal{M}$ point to the left child.

\item After we finished {\bf Triset} one time,   we will select a new polynomial set $\QS$ from $\PS^*$ at Step 3 of {\bf BCS} and run {\bf Triset} again.  At this time, we let $\mathcal{M}$ point to the node corresponding to $\QS$ and execute the operations in II) again.

\end{enumerate}

The following figure shows the zero decomposition tree of Example \ref{ex_bcs}. The root node $\PS$ is corresponding to the input polynomial system.

\

\begin{center}
{\small Figure 1. The zero decomposition tree of Example \ref{ex_bcs}}

\vspace{10pt}
\begin{tikzpicture}[level distance=2.2cm,
  level 1/.style={sibling distance=7cm},
  level 2/.style={sibling distance=3.5cm},
   level 3/.style={sibling distance=2cm},
   leaf/.style = {shape=circle, draw, solid, minimum size=1cm},
   normaledge/.style={solid},
    border/.style={dashed }
   ]

   \node[leaf]{$\PS$}
    child[normaledge,->]
    { node[leaf]{$\PS^1$}
      child
      {  node[leaf]{$\PS^2$}
         child[normaledge]{node[leaf]{$\PS^3$}
         edge from parent node[above left]{\footnotesize$I_3=1$}
       }
         child[border]{node[leaf]{$\PS_3$} edge from parent node[above right]{\footnotesize$I_3=0$}
       }
        edge from parent node[above left]{\footnotesize$I'_2=1$}
       }
      child[border]
      {
         node[leaf]{$\PS_2$}
         child[normaledge]{node[leaf]{$\PS_2^1$}
         edge from parent node[above left]{\footnotesize$I_6=1$}
       }
         child[border]{node[leaf]{$\PS_6$}
         edge from parent node[above right]{\footnotesize$I_6=0$}
       }
        edge from parent node[above right]{\footnotesize$I'_2=0$}
       }
        edge from parent node[above left]{\footnotesize$I_1=1$}
    }
    child [border,->]
    {  node[leaf]{$\PS_1$}
       child [normaledge]
       { node[leaf]{$\PS_1^1$}
         child[normaledge]{node[leaf]{$\PS_{1}^2$}
         edge from parent node[above left]{\footnotesize$I_5=1$}}
         child[border]{node[leaf]{$\PS_5$ }
          edge from parent node[above right]{\footnotesize$I_5=0$}
         }
        edge from parent node[above left]{\footnotesize$I_4=1$}}
       child [border]
       {  node[leaf]{$\PS_4$}
                 child[normaledge] { node[leaf]{$\PS_{4}^1$}
                           edge from parent node[above left]{\footnotesize$I_7=1$}
                   }
                 child[border]{node[leaf]{$\PS_{7}$}
                   edge from parent node[above right]{\footnotesize$I_7=0$}
                   }
          edge from parent node[above right]{\footnotesize$I_4=1$}
        }
       edge from parent node[above right]{\footnotesize$I_1=0$}
       };
\end{tikzpicture}

\end{center}

\

We call the external path of  a zero decomposition tree a {\em solving branch}.
It is obvious that a solving branch is corresponding to one execution of {\bf Triset}.
Moreover, in this branch there is a node which is corresponding to the input of {\bf Triset}. Obviously, when the solving branch is corresponding to first execution of  {\bf Triset}, this node is the root node. Otherwise, this node is the last node which is the right child of some node on this path.
Moreover, we call the {\em initial depth} of a solving branch to be the depth of this node. For example, for the path from $\PS$ to $\PS_2^1$ in Figure 1, its initial depth is determined by the depth of node $\PS_2$, which is equal to 2.

%
%
%
%

\

Based on the zero decomposition tree, we can estimate the complexity of {\bf BCS} by combing the complexity of {\bf Triset} and the number of solving branches. Precisely speaking, if {\bf Triset} has a complexity bound $c$ and the depth of the zero decomposition tree is bounded by $b$, then the complexity of {\bf BCS} is bounded by $c\cdot 2^b$.

\subsection{The complexity of {Triset}}

\begin{Lemma}\label{lm_com_subs}
For a polynomial system $\PS=\{f_1,f_2,\ldots,f_m\}\subset \F_2[x_1,x_2,\ldots,x_n]$ with $\tdeg(f_i)\le d$ and a linear polynomial $x_c+L$ with
class $c$.
The bit-size complexity of the operation of substituting all $x_c$ with $L$ in $\PS$ is $O(dmn^{d+2}\log(n))$.
\end{Lemma}
\begin{Proof}
A polynomial $f_i$ can be written as $P_1x_c+P_2$. Then the substitution process of this polynomial is computing $P_1L+P_2$. Since $\tdeg(f_i)\le d$, we have $\tdeg(P_1)\le d-1$ and $\tdeg(P_2)\le d$. Moreover, we know that $P_1$ and $P_2$ have at most $n-1$ variables, thus $P_1$ has
at most $\sum_{i=0}^{d-1}{n-1 \choose i}$ terms and $P_2$ has at most $\sum_{i=0}^{d}{n-1 \choose i}$ terms. $L$ is a linear polynomial with at most $n-1$ terms. For computing $P_1L$, we need to do  at most $(n-1)\sum_{i=0}^{d-1}{n-1 \choose i}$ times of monomial multiplication.  For two monomial with $n-1$ variables,  the multiplication is equal to the addition of two vectors with $n-1$ dimension, thus we need $n-1$ operations. Thus, we need $(n-1)^2\sum_{i=0}^{d-1}{n-1 \choose i}$ operations to achieve the monomials of $P_1L$. To compute $P_1L$, we need to sum up all
these monomials, and the complexity of this process is equal to the complexity of sorting all these monomials, which is $(n-1)((n-1)\sum_{i=0}^{d-1}{n-1 \choose i})\log((n-1)\sum_{i=0}^{d-1}{n-1 \choose i})=O(dn^{d+2}\log(n))$.

Note that , $P_1L$ and $P_2$ are two polynomials whose terms have been sorted, thus the complexity of $P_1L$ plus $P_2$ is $2(n-1)\sum_{i=0}^d{n-1 \choose i}=O(n^{d+1})$. Then, the complexity of computing $P_1L+P_2$ is $O(dn^{d+2}\log(n))+O(n^{d+1})=O(dn^{d+2}log(n))$. Hence, the complexity of $m$ times of substitution is $O(dmn^{d+2}\log(n))$.
\end{Proof}

\

In the following paragraphs, we define $\tdeg(\PS)$, the degree of a polynomial system $\PS$, to be the highest total degree of the elements in $\PS$.

Now we introduce the concept of backtracking for the polynomials occurring in the whole procedure of {\bf BCS}. One can find that except adding $I+1$ into the current branch and $I$ into the new generated branch, the purpose of other operations in {\bf Triset} is replacing a polynomial $P$ with another polynomial $R$. Precisely speaking, there are three kinds of operations.

\begin{itemize}[(1)]
\item Replace $P=Ix_c+U$ with $R=x_c+U$ or $U$ after zero decomposition.

\item Replace $P=x_c+U_1$ with $R=U_1+U_2$ after compute $P+P'=(x_c+U_1)+(x_c+U_2)$.

\item Replace $P=g_1x_k+g_2$ with $R=g_1L+g_2$, after substituting $x_k$ with the linear polynomial $L$ in $P$.
\end{itemize}

\noindent Therefore,  in the following, we say $R$ can {\em backtrack} to $P$, if there is a polynomial sequence $P, P^1, P^2,\ldots, P^s,R$, s.t $P^1$ replaced $P$, $P^{i+1}$ replaced $P^{i}$ and $R$ replaced $P^s$ by the above operations. Moreover, $P$ can backtrack to itself.


\begin{theorem}\label{th_triset}
Let $\PS=\{f_1,f_2,\ldots,f_m\}$ be the input polynomial system of {\bf BCS}, and
$tdeg(\PS)=d$.
If the depths of branches in the zero decomposition tree  are not greater than $b$, then the bit-size complexity of solving any branch of {\bf BCS} is bounded by  {\em $O(dm(m+b)\log(n)n^{d+3})$}.
\end{theorem}
\begin{Proof}
At first, we consider the first branch.
The major operations in {\bf Triset} are additions of two monic polynomials and substituting variables with linear polynomials. The complexity of other operations can be ignored when compared to these two kinds of operations.
First, we consider the addition of two monic polynomials. Additions may occur in two cases. The first one is in \addreduce and the second one is when we compute add-remainder.
Note that, after each time of zero decomposition, the number of polynomials in the current branch, that is the number of polynomials in $\PS\cup\MS\cup \A$, will increase by at most 1.
Hence, after $b$ times of zero decomposition, we will add $b$ polynomials into the current branch.
Then we can deduce that each polynomial occurring in $\PS\cup\MS\cup \A$ at any step of {\bf Triset} can backtrack to one of the $m$ input polynomials or one of the $b$ newly added polynomials.

For any class $c$, consider the monic polynomials involved in \addreduce. Obviously, for these polynomials, additions can only be performed within the ones that cannot backtrack to the same polynomial. It implies that for any class $c$, one can execute at most $m+b$ times of addition in \addreduce. The bit-size complexity of adding two polynomials with degree $d$ is $O(n^{d+1})$.
Therefore, the bit-size complexity of the addition operations in \addreduce is bounded by $O((m+b)n^{d+2})$.

Now we consider the complexity of computing add-remainders. Actually, we need at most $n$ times of addition for computing one add-remainder. In each time of zero decomposition,  if $\arem(I, \MS)$ is not a constant, we need compute one add-remainder, and if $\arem(I, \MS)$ is a constant, we need compute at most $d-1$ add-remainders, since $\tdeg(I)$ decrease strictly. Hence, for $b$ times of zero decomposition, we need at most $b(d-1)n$ times of addition. Therefore, the bit-size complexity of the addition operations in computing add-remainders is bounded by  $O(b(d-1)n^{d+2})$.

When we execute the substitution in \simplify one time, we will eliminate one variable, thus \simplify can be executed at most $n$ times. Each time, the number of polynomials that we need to
do substitutions is at most $m+b$.
Hence, according to Lemma \ref{lm_com_subs}, the complexity of executing \simplify $n$ times is $O(dm(m+b)n^{d+3}\log(n))$.

In summary, the bit-size complexity of solving the first branch of the zero decomposition tree is bounded by $O(n^{d+2}(m+b+bd)+O(dm(m+b)\log(n)n^{d+3})=O(dm(m+b)\log(n)n^{d+3})$. Moreover, the above proof can be easily extended to other branches, since the polynomials in other branches polynomials can also backtrack to the input polynomials and the newly added polynomials.
\end{Proof}

\

The results in the following subsection will show that  when $d$ is fixed, the value of $b$ in Theorem \ref{th_triset} is polynomial w.r.t $n$ and $m$, which means that  when $d$ is fixed, solving one branch has polynomial-time complexity.

\subsection{The complexity of {\bf BCS}}

In this section we will analyze the complexity of {\bf BCS}, and our key problem is to estimate the bound of the depth for the solving branches in the zero decomposition tree.
First,  we consider a fundamental case, that is solving quadratic Boolean polynomial systems. Quadratic Boolean polynomial systems are typical nonlinear systems, and the problem of solving them is called the Boolean MQ problem.

\begin{Lemma}\label{lm_depth_d2}
Suppose the input of {\bf BCS} is a quadratic polynomial system with $n$ variables. Then the depth of the branches for the zero decomposition tree of {\bf BCS} is less than $n$.

\end{Lemma}
\begin{Proof}
Note that, each time we generate a new branch, we do zero decomposition for one time.
We add $I+1$ in the current branch and $I$ in the new generated branch, where $\tdeg(I)< \tdeg(P)$ and $P$ is the chosen polynomial.
Since the input system is quadratic and the degree of polynomials will not increase in the whole process of {\bf BCS}, we have $\tdeg(P)=2$. Therefore, $I+1$ and $I$ are linear, which means in the current branch and the new generated branch, we both have a new linear polynomial. Then, by \simplify, we can eliminate one variable by the linear polynomial.
Therefore, this can only happen at most $n-1$ times, which means the depth of the tree is less than $n$.
\end{Proof}

\

Note that when solving quadratic system, branches with bigger initial depths have less variables, since more variables were eliminated in the former processes. Thus, the complexity of solving a branch with bigger initial depth is smaller. Based on this observation, we have the following lemma.

\begin{Lemma}\label{lm_complex_triset}
Suppose the input of {\bf BCS} is a quadratic polynomial system with $n$ variables and $m$ polynomials. Let $\PS_1$ be the system corresponding to a branch with initial depth $b_0$ and depth $b$.
 Then the bit-size complexity of {\bf Triset}$(\PS_1)$ is $O((m+b)(n-b_0+1)^{5}{\log(n-b_0+1)}))$.
\end{Lemma}

\begin{Proof}
It is obvious that  after one time of zero decomposition, the number of polynomials in this branch will increase at most by one. Therefore, at any step of {\bf Triset}, the number of polynomials in this branch is always not bigger than $m+b$.

Since the initial depth of this branch is $b_0$, we have already eliminated $b_0-1$ variables before solving this branch. Thus, in the \simplify process, we can do the substitution $n-b_0+1$ times, and the complexity
is $O((m+b)(n-b_0+1)^{5}{\log(n-b_0+1)})$ according to Lemma \ref{lm_com_subs}.

Now let us estimate the complexity of addition operations. The complexity of adding two quadratic polynomials with $n-b_0$ variables is $O((n-b_0)^3)$. For each class, the number of polynomials is not bigger  than $m+b$, and the number of different classes is at most $n-b_0$.  Therefore the number of additions is not bigger than $(n-b_0)(m+b)$, and the complexity of addition operations is $O((n-b_0)^4(m+b))$.
Then, the complexity of {\bf Triset} is $O((m+b)(n-b_0+1)^{5}{\log(n-b_0+1)}))+O((n-b_0)^4(m+b))=O((m+b)(n-b_0+1)^{5}{\log(n-b_0+1)}))$.
\end{Proof}

\begin{Lemma}\label{lm_com_r2}
Let $\PS$ be a quadratic polynomial system with $n$ variables and $m$ polynomials. If the depth of the branches of {\bf BCS}$(\PS)$
are not bigger than $b$, then the bit-size complexity of {\bf BCS}$(\PS)$ is bounded by
\begin{enumerate}[1)]
\item $O((b+1)(m+b)2^{b-1}(n-b+1)^6)$, when $b<n-9.66$ .
\item  $O((b+1)(m+b)2^{n})$, when $b\ge n-9.66$.
\end{enumerate}

\end{Lemma}
\begin{Proof}
For a solving branch with depth $b$, except the root node there are $b$ nodes in this branch. They can form a sequence
$\{E_1,E_2,\ldots,E_b\}$, where $E_i=L$ or $R$, which means the i-th node is a left child or a right child respectively.
For example, the node sequence of the first branch is $\{L,L,\ldots,L\}$.
Then, the node sequence for a branch with initial depth $b_0>0$, must have the form
$$\{E_1,E_2,\ldots,E_{b_0-1},R,\underbrace{L,L,\ldots,L}_{b-b_0}\},$$
where $E_i, 1\le i \le {b_0-1}$ can be either $L$ or $R$.
Thus, the number of branches with initial depth $b_0$ is at most $2^{b_0-1}$.
According to Lemma \ref{lm_complex_triset},  the bit-size complexity of {\bf Triset}  is $O((m+b)(n-b_0+1)^{5}{\log(n-b_0+1)}))\le O((m+b)(n-b_0+1)^{6}$. Note that, there is only one branch with initial depth $0$, and the complexity of solving this branch is bounded by $O((m+b)n^6)$.
Then, the complexity of {\bf BCS} is bounded by $(m+b)(n^6+\sum_{k=1}^{b}2^{k-1}(n-k+1)^{6})$.
Function $2^{x-1}(n-x+1)^6$ reaches its maximal value when $x=n-6/\log2+1=n-9.66$. Thus,
if $b<n-9.66$,
 $(m+b)(n^6+\sum_{i=1}^{b}2^{k-1}m(n-k+1)^{6}))\le O((m+b)(b+1)2^{b-1}(n-b+1)^6)$.
 If $b\ge n-9.66$, $(m+b)(n^6+\sum_{k=1}^{b}2^{k-1}(n-k+1)^{6}\le (m+b)(b+1)2^{n-9.66}(9.66)^8=O((b+1)(m+b)2^{n})$
\end{Proof}

\

Based on Lemma \ref{lm_depth_d2} and \ref{lm_com_r2}, we have the following theorem.

\begin{theorem}
Let $\PS$ be a  quadratic polynomial system with $n$ variables and $m$ polynomials.
The bit-size complexity of {\bf BCS}$(\PS)$ is bounded by
$O(n(m+n)2^{n})$.
\end{theorem}

The  bound $O(n(m+n)2^n)$ is the complexity bound of {\bf BCS} in the worst case.
To the best of the authors' knowledge, the best complexity result for solving Boolean quadratic polynomial system without side conditions is $4\log_2(n)2^n$ bit operations for the fast exhaustive search method proposed in \citep{fast_search}.
When some assumption is made for the system, the complexity of solving Boolean MQ problem can be less than $O(2^n)$. In \citep{bardet_comp}, Bardet et al. proposed an algorithm by combining  exhaustive search and  spare linear algebra, the complexity of this algorithm is $O(2^{0.841n})$.
when $m=n$ under some precise algebraic assumptions which are satisfied with probability very close to 1. Moreover, a probabilistic variant of their algorithm (Las Vegas type) has expected complexity $O(2^{0.792n})$.

\

In the following, we consider the polynomial systems with degree higher than $2$.

\begin{Proposition}\label{th_complexity1}
Let $\PS=\{f_1,f_2,\ldots,f_m\}$ be the input polynomial system of {\bf BCS}, and
$tdeg(\PS)=d$.
For any {\em \texttt{Choose}}, the depths of the branches in the zero decomposition tree are not bigger than$m\sum_{j=1}^{d-1}{n \choose j}$.
\end{Proposition}

\begin{Proof}
It is sufficient to prove the theorem in the worst case, that is the input polynomials are all with class $n$, and when we choose a polynomial $P=Ix_c+U$ to do zero decomposition, the polynomial we add into $\PS$ and  the polynomial replacing $P$ in $\PS$ are all with class $c-1$.

We consider the first branch. Let $M_c$ denote the set of polynomials with class $c$ that can be backtracked by other polynomials. Moreover, if there are several polynomials with $c$ which can be backtracked from the same polynomial with lower class, then only one of them, as a canonical element,  is in $M_c$.
Obviously $|M_n|\le m$, and the number of zero decomposition  for polynomials with class $n$  is not larger than $m$. Then, according to the proof of Theorem \ref{th_triset}, polynomials in $M_{n-1}$ can be grouped into the following two polynomial  sets:

\begin{enumerate}
\item $\PS_{I}$: the newly added polynomials.

\item $\PS_{R}$: the polynomials which can backtrack to polynomials with class $n$.

\end{enumerate}

\noindent Obviously, we have $|\PS_{I}| \le M_n \le m$, $\tdeg(\PS_{I})\le d-1$, $|\PS_{R}| \le M_n \le m$ and $\tdeg(\PS_{R})\le d$.

Similarly, the polynomials in $M_{n-2}$ can be sorted into the following four polynomial sets:

\begin{enumerate}

\item $\PS_{II}$:  the new polynomials which are added when we choose polynomials in $\PS_{I}$ to do zero decomposition.

\item $\PS_{IR}$: the new polynomials which are added when we  choose polynomials in $\PS_{R}$ to do zero decomposition.

\item $\PS_{RI}$: the polynomials which can backtrack to polynomials in $\PS_I$.

\item $\PS_{RR}$: the polynomials which can  backtrack to polynomials in $\PS_R$.

\end{enumerate}

Then, we have $|\PS_{JJ}|,|\PS_{RR}|,|\PS_{JR}|,|\PS_{RJ}|\le m$. $\tdeg(\PS_{JJ})\le d-2$, $\tdeg(\PS_{JR})\le d-1$, $\tdeg(\PS_{RJ})\le d-1$ and $\tdeg(\PS_{RR})\le d$.

Recursively, we can define ${\PS_{O_1O_2\cdots O_{k}}}$, where $O_i$ is $I$ or $R$, and the polynomials in $M_{n-k}$ can be sorted into these polynomial sets. We have
$|\PS_{O_1O_2\cdots O_{k}}|\le m$ and
$\tdeg(\PS_{O_1O_2\cdots O_{k}})\le d-s$, where $s$ is number of $I$ in these subscripts $O_i$.
When we choose a polynomial to do zero decomposition, its total degree must be higher than 1. Therefore, for class $n-k$, when we do zero decomposition, we can only choose the polynomials in such $\PS_ {O_1O_2\cdots O_{k}}$ that the number of $I$ occurring in $O_1, O_2,\ldots, O_k$ is at most $d-2$. It means that the number of zero decompositions for class $n-k$ is not larger than $m(\sum_{i=0}^{d-2}{k \choose i})$
Hence, the total number of zero decompositions is bounded by $m\sum_{k=0}^{n-1}(\sum_{i=0}^{d-2}{k \choose i})=m\sum_{j=1}^{d-1}{n\choose j}$.

It is easy to see that the above bound can be easily extended to other branches, since we only need the property that when each time we do zero decomposition only one new polynomial with degree lower than the degree of the chosen polynomial is added into the current branch,  which is satisfied for any branches.
\end{Proof}

\begin{remark}
Note that the properties we used in the proof of Proposition \ref{th_complexity1} are also valid for the {\bf MFCS} algorithm proposed in \citep{cs_fq}. Hence this depth bound is also valid for {\bf MFCS}.

\end{remark}

The above proposition shows that by any choose function, the depth of the zero decomposition tree is bounded by $m\sum_{j=1}^{d-1}{n \choose j}$. Actually,  the depth can be much smaller, when we use some specific \fchoose functions.
In the following, we consider the following choose function.

\begin{center}
$\fchoose_1$: Choose a polynomial $P$ from a polynomial set $\PS$ s.t. $\tdeg(P)=min_{f\in \PS}\tdeg(f).$
\end{center}

\noindent Now we estimate the complexity of {\bf BCS} with $\fchoose_1$ as the choose function. First, we prove the following lemma.

\begin{Lemma}\label{lm_com_n}
Let $\PS$ be a polynomial system with $n$ variables. Suppose in some step of {\bf Triset}$(\PS)$, the polynomial set $\A\cup \MS$ has $n$ elements with different classes, then {\bf Triset} will terminate without doing zero decomposition in the following steps.
\end{Lemma}
\begin{Proof}
From the definition of $\A$ and $\MS$, we know that $\A$ and $\MS$ contain some monic polynomials. Note that in {\bf Triset}, we execute \simplify after we generated a new linear polynomial. From the assumption, we know that the classes of the elements in $\MS$ are different from those in $\A$, hence $\A \cup \MS$ forms a monic triangular set. Obviously, the element in $\A \cup \MS$ with the lowest class will have the form $x_1+b$, where $b=0$ or $1$. Then, after \simplify the element with class $2$ will also have the form $x_2+b$, where $b=0$ or $1$. Recursively, all the elements will have the form $x_c+b$ after \simplify, which means the values of the variables are fixed, then other polynomials in $\PS$ will be converted into constant after \simplify, hence no more zero decomposition is needed.
\end{Proof}

\begin{Proposition} \label{th_depth}
Let $\PS=\{f_1, f_2, \ldots, f_m\}\subset \F_2[x_1,x_2,\ldots,x_n]$ be a polynomial system with degree $d$,  and $\fchoose_1$ be the choose function of {\bf Triset}, then {\bf Triset}$(\PS)$ will terminate after $(2d-3)n$ times of zero decomposition.
\end{Proposition}
\begin{Proof}
For the polynomial sets  $\PS$, $\MS$, $\A$ at any step of  {\bf Triset}, we can define an index vector $\mathcal{T}=(\bf{d_0, d_1, \ldots, d_n})$ as follows.
For $\d_0$, we have:
\begin{enumerate}[1)]
\item If there is a linear polynomial in $\PS$,  set $\d_0$ to be 1.

\item If there are non-monic polynomials in $\PS$, set $\d_0$ to be the lowest total degree for these non-monic polynomials.

\item If all polynomials in $\PS$ are monic, set $\d_0$ to be $d$. If $\PS=\emptyset$, set $\d_0$ to be $0$.

\end{enumerate}
For $\d_i, 1\leq i \le n$,  we have:
\begin{enumerate}[1)]
\item If the classes of monic polynomials in $\A\cup \MS$ are not equal to $i$, then set $\d_i$ to be $d$.

\item If there is a linear polynomial with class $i$ in $\A$ , then set $\d_i$ to be $1$.

\item If there is a monic polynomial $M$ with class $i$ in $\MS $, we set $\d_i$ to be $\tdeg(M)$. 
\end{enumerate}

In the following, we show that after zero decomposition, the sum of all entries in $\T$, denoted by $Sum(\T)$,  will strictly decrease. Note that at Step 14 the polynomials in $\PS$ are not monic. Hence $\d_0=\tdeg(P)$, where $P$ is the chosen polynomial. Suppose $\initial(P)=I$. Then, there are five cases.

\begin{enumerate}[1)]

\item $I$ is not monic, hence $\arem(I, \MS)=I$.  Since $\tdeg(I) < \tdeg(P)$, and $P$ is the non-monic polynomial in $\PS$ with lowest degree, then after adding $I+1$ into $\PS$, the value of $d_0$ will  at least decrease by $1$. Moreover, the value of other elements in $\T$ will not increase. Thus $Sum(T)$ will strictly decrease.

\item $I$ is linear, then $\arem(I, \MS)=I$. Obviously, $d_0$ is equal to $1$ after adding $I+1$ into $\PS$, and the value of other elements in $\mathcal{T}$ will not increase. Thus $Sum(\mathcal{T})$ will strictly decrease.

\item $I$ is monic, nonlinear, and $\arem(I, M)=I$. It means that  the classes of the polynomials in $\MS$ are not equal to $c=\cls(I)$, hence before zero decomposition, $\d_c$ is $d$.  It implies that before zero decomposition, $\d_0+\d_c= \tdeg(P)+d$.  Since $\tdeg(I)<\tdeg(P)$, then after adding $I+1$ into $\MS$, we have $\d_0\le d$, $\d_c= \tdeg(I+1)\le \tdeg(P)-1$, which implies $\d_0+\d_c\le d+\tdeg(P)-1$. Moreover, the value of other elements in $\T$ will not increase, hence $Sum(\mathcal{T})$ will strictly decrease.

\item $\arem(I, \MS)\ne I$ and $\arem(I, \MS)$ is not a constant. It means that $I$ is monic, and there is some polynomials $M_1, M_2, \ldots, M_k$ in $\MS$ such that $I+M_1+M_2+\cdots+M_k=\arem(I, \MS)$. Suppose the polynomials in $\rseq(I, \MS)$ are $M'_1, M'_2, \ldots, M'_k$, and $\cls(M_i)=\cls(M'_i)=c_i$. Note that, for two polynomials $P$ and $Q$, $\tdeg(P)+\tdeg(Q)\ge \tdeg(P+Q)+\tdeg(Q)$, if $\tdeg(Q)\le \tdeg(P)$. Hence, we can deduce that that
$\tdeg(\arem(I,\MS))+\tdeg(M'_1)+\cdots+\tdeg(M'_k)\le \tdeg(I)+\tdeg(M_1)+\cdots+\tdeg(M_k)<
\tdeg(P)+\tdeg(M_1)+\cdots+\tdeg(M_k)$. If $R=\arem(I,\MS)$ is monic and nonlinear, suppose $\cls(R)=c$, then we can deduce that the value of $\d_0+\d_c+\d_{c_1}+\cdots+\d_{c_k}$ decrease strictly.  Otherwise, it is easy to see that the value of $\d_0+\d_{c_1}+\cdots+\d_{c_k}$ decrease strictly. It implies that $Sum(\mathcal{T})$ will strictly decrease.

\item $\arem(I, \MS)$ is a constant. By Step 25-27, we generate a  new polynomial $I'$ such that $\tdeg(I')<\tdeg(I)$ and $\arem(I', \MS)$ is not a constant. Then we set $I$ to be $I'$, add $\arem(I, \MS)$ into $\PS$ and update $\MS$ by $\rseq(I, \MS)$. It is obvious that one of the above four cases will occurs, thus $Sum(\mathcal{T})$ will strictly decrease.

\end{enumerate}

In all, for any cases, $Sum(\mathcal{T})$ will strictly decrease after zero decomposition.
Now we consider the variation of $Sum(\mathcal{T})$ after \addreduce and \simplify.
In \addreduce, we compute the addition of two nonlinear and monic polynomials $Q_1$ and $Q_2$, where $\cls(Q_1)=\cls(Q_2)=c$, and keep the one with lowest degree in $\MS$. From the definition of $\d_i$, we know that $\d_c$ will not increase after addition, and obviously $\d_0$ will not increase. Therefore, we can conclude that after \addreduce, $Sum(\mathcal{T})$ will not increase.

Now we consider \simplify.
In \simplify, Loop 3 may iterate several times. We focus on the first time.
Since there are several linear polynomials in $\PS$,  we have $\d_0=1$.  We consider the one with lowest class in these linear polynomials, and denote it by $L=x_c+l$, where $c$ is the class of $L$. Then, there  are two cases.
\begin{enumerate}[1)]

\item  Before \simplify, the classes of polynomials in $\MS$ are not equal to $c$, then $\d_c=d, \d_0=1$. After substituting $x_c$ with $l$ and add $L$ into $\A$, we have $\d_c=1, \d_0\le d$. It is easy to see that after each iteration of Loop 3, $\d_i$ will not increase. Hence, $Sum(T)$ doesn't increase after \simplify.

\item  Before \simplify, there is a polynomial $Q$ with degree $d'$ in $\MS$ such that $\cls(Q)=\cls(L)=c$, then $\d_c+\d_0=1+d'$. After substituting $x_c$ with $l$, we will achieve a nonlinear polynomial $Q+L$ with $\tdeg(Q+L)\le d'$.

\begin{itemize}

\item If $Q+L$ is not monic, then $\d_0\le d'$ after adding $Q+L$ into $\PS$. Note that $\cls(Q+L)<\cls(L)$, thus the classes of other linear polynomials occurring in \simplify will be bigger than $\cls(Q+L)$, which means $Q+L$ will not be changed in the following substitutions. Hence, after \simplify, $Q+L$ is still in $\PS$, then $\d_0\le d'$, $\d_c = 1$. Therefore, $Sum(\T)$ doesn't increase after \simplify.

\item If $Q+L$ is a monic polynomial,  and  before \simplify the polynomials in $\MS$ have classes different from $c'=\cls(Q+L)$, then we have $\d_{c'}=d, \d_0=1, \d_c=d'$ before \simplify. Since $Q+L$ is a monic polynomial and it will not be changed by the following substitutions,  this polynomial or another polynomial with same class and lower total degree will be added into $\MS$ after the following \addreduce process. Therefore, after \simplify and the following \addreduce, $\d_0\le d, \d_c=1, \d_{c'}\le d'$, which means $Sum(\T)$ will not increase.

\item If $Q+L$ is a monic polynomial and $c'=\cls(Q+L) = \cls(M_j)$ for some $M_j\in \MS$, then $Q+L$ may becomes $0$ after the following \addreduce. In this case, after \simplify and the following \addreduce, $\d_0$ may become $d$, and $\d_{c'}$ may not decrease, which means $Sum(T)$ may increase. In the worst cases that $\d_c=2$ before \simplify, $Sum(\mathcal{T})$ will increases at most $(d+1)-(1+2)=d-2$.
\end{itemize}

\end{enumerate}

We know that when $\T=(d,d,\ldots,d)$, $Sum(\mathcal{T})$ reaches its maximal value $d(n+1)$.
Suppose we have executed zero decomposition for $(2d-3)n$ times.
Note that for $Sum(\T)$, the increase cases only happen after \simplify was executed, thus it can occur  at most  $n$ times. It means that $Sum(\mathcal{T})$ can increase at most by $n(d-2)$. Thus, after  $(d-1)n+(d-2)n$ times of zero decomposition, we have $Sum(\mathcal{T})\le d(n+1)+n(d-2)-(d-1)n-(d-2)n=n+d$.
Now we show that when $Sum(\mathcal{T})= d+n$, {\bf Triset} will terminate without further zero decomposition. There are two cases for $Sum(\mathcal{T})=d+n$. The first case is that $\d_1, \d_2, \ldots,\d_n<d$, which means for any $1\le i\le n$,  there is a monic polynomial with class $i$  in $\A\cup \MS$. According to Lemma \ref{lm_com_n},  {\bf Triset} will terminate without further zero decomposition.  The second case is that $\d_j=d$ for some $1\le j\le n$. Since $Sum(\mathcal{T})= d+n$, we have  $\d_0=1, \d_1=1,\ldots, \d_{j-1}=1,\d_{j}=1, \ldots, \d_{n}=1$. Obviously, $x_1, x_2, \ldots, x_{j-1}, x_{j+1},\ldots, x_{n}$ will not occur in  polynomials in $\PS$. Hence the polynomials in $\PS$ with lowest degree have degree $1$ and have leading variable $x_j$, which implies {\bf Triset} will terminate without further zero decomposition. It is easy to see that if $Sum(\mathcal{T})< d+n+1$,  {\bf Triset} will  also terminate without further zero decomposition.
\end{Proof}

\

The above proposition shows that the first branch of the zero decomposition tree has depth not bigger than $(2d-3)n$. Note that, in the above proof, the critical property we used is $\tdeg(I+1)< \tdeg(P)$. For other branches generated by considering $I=0$, we also have $\tdeg(I)=\tdeg(I+1)<\tdeg(P)$, hence same result can be proved. Consequently, we have the following theorem.

\begin{theorem}\label{th_depth}
Let $\PS=\{f_1, f_2, \ldots, f_m\}\subset \F_2[x_1,x_2,\ldots,x_n]$ with $\tdeg(\PS)= d$ be the input of {\bf BCS}, and $\fchoose_1$ be the choose function. Then the depths of  the branches of the zero decomposition tree are not bigger than $(2d-3)n$.
\end{theorem}

By combining Theorem \ref{th_triset} and \ref{th_depth}, we have the following complexity bound about {\bf BCS}.

\begin{theorem}\label{th_bcs_comp1}
Let $\PS=\{f_1, f_2, \ldots, f_m\}\subset \F_2[x_1,x_2,\ldots,x_n]$ with $\tdeg(\PS)= d$ be the input of {\bf BCS}, and $\fchoose_1$ be  the choose function. Then the complexity of {\bf BCS} is bounded by  $O(d(m^2+(2d-3)nm)\log(n)n^{d+3}2^{(2d-3)n})$.
\end{theorem}

\

Furthermore,  we consider the following choose function.


$\fchoose_2$: Choose a polynomial $P$ from a polynomial set $\PS$ such that $\tdeg(\initial(P))=min_{f\in \PS}\tdeg(\initial(f))$

The following theorem shows that for $\fchoose_2$ the above complexity bound for {\bf BCS} is still valid.

\begin{theorem}\label{th_bcs_comp2}
Let $\PS=\{f_1, f_2, \ldots, f_m\}\subset \F_2[x_1,x_2,\ldots,x_n]$ with $\tdeg(\PS)= d$ be the input of {\bf BCS}, and $\fchoose_2$ be the choose function. Then the complexity of {\bf BCS} is bounded by  $O(d(m^2+(2d-3)nm)\log(n)n^{d+3}2^{(2d-3)n})$.
\end{theorem}

\begin{Proof}
It is sufficient to show that the depth of the zero decomposition tree is not larger than $(2d-3)n$.
First, we consider the first branch. We define an new index vector $\mathcal{T}'=(\d_0, \d_1, \ldots, \d_n)$  for the polynomial sets $\PS$, $\MS$, $\A$ in {\bf Triset},  where the definitions of $\d_1, \ldots, \d_n$ are same as those of $\mathcal{T}$ in the proof of Proposition \ref{th_depth}, and $\d_0$ is defined as follows.
\begin{itemize}

\item If there is a linear polynomial in $\PS$,  set $\d_0$ to be 0.

\item If there are non-monic polynomials in $\PS$, set $\d_0$ to be the lowest total degree of the initials of these non-monic polynomials.

\item If all polynomials in $\PS$ are monic, set $\d_0$ to be $d-1$. If $\PS=\emptyset$, set $\d_0$ to be $0$.
\end{itemize}

Similarly, $Sum(\mathcal{T}')$ is defined to be the sum of the entries in $\mathcal{T}'$. Obviously, we can prove that after different operations the variation of $Sum(T')$ is same as that of $Sum(T)$. Consider the maximal possible value of $Sum(\mathcal{T}')$ which is equal to $(n+1)d-1$ and achieved when $\mathcal{T}'=\{d-1,d,d,\ldots,d\}$.
Then, after do zero decomposition for $(2d-3)n$ times, $Sum(T')$ is at most $d+n-1$. Then, we have either $\d_1, \d_2, \ldots, \d_n<d$, or $\d_0=0, \d_k=d, \d_i=1$, for some $1\le k \le n$ and any $1\le i \le n, i\ne k$. For these two cases, we can deduce that {\bf Triset} will end without further zero decomposition.
This implies that the depth of the first branch is not larger than $(2d-3)n$. Moreover, it is easy to see that this depth bound is still valid for other branches.
\end{Proof}

\begin{remark}
For a polynomial set $\PS$, we have $min_{f\in \PS}{\tdeg(\initial(f))} \le min_{f\in \PS}{\tdeg(f)}$. It means that compared with choosing the polynomial with lowest total degree,  choosing the polynomial whose initial has lowest total degree may make $\d_0$ decrease faster, and this can induce a lower experimental complexity.
\end{remark}

\subsection{A variant algorithm with lower complexity bound}

In this section, we show that by slightly modifying {\bf BCS}, the case of $Sum(\T)$(or $Sum(\T')$) increasing after \simplify can be absolutely avoided.

We perform the following modification on {\bf BCS}, and name the new algorithm {\bf BCS$_g$}:

Suppose $\MS$ is changed after zero decomposition, \addreduce, or \simplify.  We sort $\MS$, such that the former elements have the lower classes,  and suppose $\MS =\{M_1, M_2,\ldots, M_t\}$ after sorting. Then, sort the different monomials in $\MS$ with respect to a graded order, and  set the different nonlinear monomials as different new variables $y_1, y_2,\ldots, y_k$. Then each polynomial $M_i\in \MS$ is a linear polynomial in variables  $y_1,\ldots, y_k$ and $x_1, \ldots, x_n$. We generate the coefficient matrix $\mathcal{M}$ about $\MS$, such that
$\mathcal{M}(y_1,\ldots, y_k, x_1,\ldots,x_n)^T=(M_1, M_2, \ldots, M_t)^T$.
Perform one-direction Gaussian elimination on $\mathcal{M}$, that is we only use the upper rows to eliminate the lower rows, and don't swap the position of two rows.
Suppose, after Gaussian elimination, $\mathcal{M}$ becomes $\mathcal{M'}$. Then set $\MS$ to be $\mathcal{M'}(y_1,\ldots, y_k, x_1,\ldots,x_n)^T$. If there are linear polynomials in $\MS$, then repeat the following operations until there is no linear polynomial in $\MS$:

 Step 1. For each linear polynomial $L= x_c+l\in \MS$, move it from $\MS$ into $\A$;

 Step 2. Substitute $x_c$ with $l$ for polynomials in $\PS\cup \MS$.

\

In the following we estimate the complexity of {\bf BCS}$_g$. First, we analyze the variation of $Sum(T)$ after zero decomposition.
Evidently, for $\mathcal{M}$, Gaussian elimination does not increase the total degree of the polynomial corresponding to each row,  since the monomials are sorted with decreasing degree.
Hence, if no linear polynomial is generated after Gaussian elimination, $\d_0$ will not changed and $\d_1, \ldots, \d_n$ will not increase, thus $Sum(\mathcal{T})$ will not increase.
Hence, if not linear polynomial is generated after Gaussian elimination, $Sum(\T)$ will strictly decrease.
In the following, we consider the case that new linear polynomials are generated after Gaussian elimination.
\begin{itemize}
\item Suppose we add a non-monic polynomial $\arem(I,\MS)$ into $\PS$ after zero decomposition, then $\MS$ may be updated by $\rseq(I, \MS)$, after computing $\arem(I, \MS)$. Since $\cls(\arem(I,\MS))$ is smaller than the classes of polynomials in $\rseq(I, \MS)$, hence is smaller than the classes of the new generated linear polynomials by Gaussian elimination. Thus, $\arem(I,\MS)$ will not be changed by substitutions w.r.t. these new linear polynomials, thus $Sum(\T)$ will strictly decrease in this case.
\item Suppose we add a monic polynomial in $\MS$ after zero decomposition. Note that,  in the proof of Proposition \ref{th_depth}, we assume that $\d_0\le d$ after zero decomposition, and this still holds after substitutions were executed, hence $Sum(\T)$ will strictly decrease.
\end{itemize}

Now we consider \addreduce. After \addreduce, suppose there is a monic polynomial $M$ in $\MS$. If  before \addreduce, there is no polynomial with class equal to $\cls(M)$ in $\MS$, we call this $M$ a {\em totally new} polynomial in $\MS$. If before \addreduce, there is a polynomial $M'$ with $\cls(M')=\cls(M)$ and $\tdeg(M')>\tdeg(M)$, then we say  $M'$ is replaced by $M$  in $\MS$.
Similarly as above,  if no linear polynomials is generated by Gaussian elimination, $\d_0$ will not  be changed and $\d_i$ with $1\le i \le n$ will not increase, hence $Sum(\T)$ will not increase.
Now suppose there are several linear polynomials generated by Gaussian elimination, and $P_0$ is the one with lowest class, where $\cls(P_0)=c$.
For $\MS$ and $\PS$ before \addreduce, let $R_1 = \{ P: P \in \MS, \cls(P)\le c\}\cup \{P: P \in \PS, \cls(P)<c\}$, and $S_1=\{\lm(f): f\in R_1\}$.
For $\MS$ and $\PS$ after Gaussian elimination, let $R_2 = \{ P: P \in \MS, \cls(P)\le c\}\cup \{P: P \in \PS, \cls(P)<c\}$, and $S_2=\{\lm(f): f\in R_2\}$.
Here $\lm(f)$ is the leading monomial of $f$ w.r.t. the graded order used in the Gaussian elimination process. Note that,  in \addreduce or Gaussian elimination, after we compute the addition of two polynomials $f$ and $g$ with $\lm(f)\le \lm(g)$, $f$ is not changed and $g$ is converted into $g'=f+g$,  hence $\{\lm(g), \lm(f)\}\subset \{\lm(g'), \lm(f)\}$.  Therefore, we can deduce that $S_1\subset S_2$. Then there are two cases.
\begin{enumerate}[1)]

\item Before \addreduce, there is no polynomial with class $c$ in $\MS$. Then before \addreduce, we have $\d_c =d, \d_0\ge 2$. After Gaussian elimination and the following substitutions, we have $\d_c=1, \d_0\le d$. Since other $\d_i$ will not increase, we have $Sum(\T)$ will decrease.

\item Before \addreduce, there is a polynomial $Q_0$ with class $c$ in $\MS$. Since after Gaussian elimination, $\lm(P_0)\ne \lm(Q_0)$ and $\lm(Q_0)\in S_2$, then $\lm(Q_0)$ must be the leading monomial of some polynomial $P_1$ in $R_2\setminus \{P_0\}$.
Then there are three case for $P_1$. The first case is  that $P_1$ is a totally new polynomial in $\MS$. The second case is that $P_1$ is in $\PS$, and it is not monic since the polynomials in $\PS$ are not monic after \addreduce. The third case is that a polynomial $Q_1$ is replaced by $P_1$ in $\MS$.
Since the leading monomials of the polynomials in $\MS$ before \addreduce are different, we have $\lm(P_1) =\lm(Q_0)\ne \lm(Q_1)$. Then $\lm(Q_1)$ must be the leading monomial of some polynomial $P_2$ in $R_2 \setminus\{P_0, Q_1\}$, and one of the above three case will happen again. Obvious, the third case can happen finite times. Hence, we can obtain a sequence $Q_0, P_1, Q_1, P_2, Q_2,\ldots,P_k$, such that $Q_i$ is replaced by $P_i$ in $\MS$, $\lm(Q_i)=\lm(P_{i+1})$, and $P_k$ is either non-monic or a totally new polynomial in $\MS$.
\begin{enumerate}
\item If $P_k$ is a non-monic polynomial, then after Gaussian elimination, we have $\d_0+\d_c+\d_{c_1}+\d_{c_2}+\cdots+\d_{c_{k-1}} \le \tdeg(P_k)+1+\tdeg(P_1)+\tdeg(P_2)+\cdots+\tdeg(P_{k-1})=D_0$, where $c_i=\cls(P_i)$.
Moreover, before \addreduce we have  $\d_0+\d_c+\d_{c_1}+\d_{c_2}+\cdots+\d_{c_{k-1}} \ge 2 +\tdeg(Q_0)+\tdeg(Q_1)+\tdeg(Q_2)+\cdots+\tdeg(Q_{k-1})=D_1$. Since $\lm(Q_i)=\lm(P_{i+1})$, we have $\tdeg(Q_i)=\tdeg(P_{i+1})$, then $D_1-D_0=1$. Note that, the new linear polynomials generated by Gaussian elimination have classes bigger than $c$, hence $P_k$ will not be changed after substitutions, which implies that $Sum(\T)$ will decrease.

\item If $P_k$ is a totally new polynomial in $\MS$, then $\d_0+\d_c+\d_{c_1}+\d_{c_2}+\cdots+\d_{c_{k-1}}+\d_{c_k}\le d+1+\tdeg(P_1)+\tdeg(P_2)+\cdots+\tdeg(P_{k-1})+\tdeg(P_k)=D_0$ after Gaussian elimination. Before \addreduce, we have  $\d_0+\d_c+\d_{c_1}+\d_{c_2}+\cdots+\d_{c_{k-1}}+\d_{c_k} \ge 2+\tdeg(Q_0)+\tdeg(Q_1)+\tdeg(Q_2)+\cdots+\tdeg(Q_{k-1})+d=D_1$. Similarly as a), we have $D_1-D_0=1$. Moreover, after the following substitutions, we still have $\d_0\le d$, thus $Sum(\T)$ will decrease.

\end{enumerate}
\end{enumerate}

Now we consider \simplify.
As proof of Proposition \ref{th_depth}, we focus on the linear polynomial with lowest class in the first iteration of Loop 3.
Suppose this linear polynomial is $L=x_c+l$, where $c=\cls(L)$. Obviously, if $\d_c=d$ before \simplify, we can prove that $Sum(\T)$ decreases.
In the following, consider the case that $\d_c<d$. Suppose there is a polynomial $P$ with class $c$ in $\MS$ before \simplify.
If $P+L$ is a non-monic polynomial, it will not be changed by the following substitutions. Hence, $\d_0\le \tdeg(P), \d_c=1$ after \simplify, and $\d_0=1, \d_c=\tdeg(P)$ before \simplify, which means $Sum(\T)$ will not increase.
If $P+L$ is a monic polynomial, then it will be in $\PS$ after the following substitution, and we will execute \addreduce in the following steps of {\bf Triset}. Let $T_0$ be the value of $Sum(\T)$ before \simplify, $T_1$ be the value of $Sum(\T)$ after \simplify and the following substitutions, and $T_2$ be the value of $Sum(\T)$ after \addreduce and the following substitutions. We will prove  $T_2\le T_0$.
\begin{itemize}

\item Suppose we obtain a linear polynomial in the Gaussian elimination after \addreduce. In this case, the above analysis about \addreduce shows that $T_2< T'$, where $T'=2+d_1+d_2+\cdots+d_n$ and $d_i$ is the value of $\d_i$ before \addreduce. Obviously, $T_1$ is equal to  $k+d_1+d_2+\cdots+d_n$, where $k\ge 2$ is the value of $\d_0$ after \simplify and the following substitutions. Hence, $T_1\ge T'>T_2$. Moreover, if $k\le \tdeg(P)$, similarly as  the case that $P+L$ is non-monic, we can prove that $T_0\ge T_1$. Since $\tdeg(P)\ge 2$, we have $T_0\ge T'$. Consequently, $T_0\ge T'>T_2$.

\item Suppose we don't obtain linear polynomials in the Gaussian elimination after \addreduce.
Then as case 2) in the above analysis about \addreduce, $\lm(P+L)$ will be the leading monomial of a polynomial $Q_0$ in $\PS$ or $\MS$ after \addreduce and the following Gaussian elimination. Then, there are three cases: $Q_0$ is a non-monic polynomial; $Q_0$ is a totally new polynomial in $\MS$; a polynomial $P_1$ is replaced by $Q_0$ in $\MS$.  Similarly as case 2) in the above analysis about \addreduce , we can prove that $T_2\le T_0$.

\end{itemize}

In summary, we proved that $Sum(\T)$ will decrease strictly after zero decomposition, and will not increase after $\addreduce$ and $\simplify$. It means that the depth of the zero decomposition tree is bounded by $(d-1)n$.

Now we consider the complexity of solving one branch for {\bf BCS}$_g$. Obviously, the difference of {\bf BCS} and {\bf BCS}$_g$ is the process of Gaussian elimination.
We show that compared to the complexity of other operations in {\bf Triset}, the complexity of Gaussian elimination is ignorable.  We know that \simplify can be executed at most $n$ times, and zero decomposition can be executed at most $(d-1)n$ times. Moreover, \addreduce can happen after zero decomposition or \simplify, hence can be executed at most $dn$ times. Thus, Gaussian elimination can be executed at most $2dn$ times. The complexity of Gaussian elimination for $n$ vectors with dimension $\sum_{i=2}^d{n \choose d}$ is bounded by $O(n^{d+2})$.
Hence for {\bf BCS}$_g$ with $\fchoose_1$ or $\fchoose_2$, the complexity of solving one branch is bounded by
$O(d(m^2+(d-1)nm)\log(n)n^{d+3})+O(2dn^{d+3})=O((dm^2+d^2nm)\log(n)n^{d+3})$.
Then, we have the following theorem.

\begin{theorem}\label{th_bcsg_comp}
Let $\PS=\{f_1, f_2, \ldots, f_m\}\subset \F_2[x_1,x_2,\ldots,x_n]$ with $\tdeg(\PS)= d$ be the input of {\bf BCS$_g$}, and set  the choose function to be $\fchoose_1$  or $\fchoose_2$.  Then the complexity of {\bf BCS$_g$} is bounded by  $O((dm^2+d^2nm)\log(n)n^{d+3}2^{(d-1)n})$.
\end{theorem}

\begin{remark} The probability of $Sum(\T)$ increasing after \simplify is very low, and we didn't observe this case happened when solving the polynomial systems in our experiments. It means that when consider experimental complexity, executing Gaussian elimination for $\MS$ is redundant, and {\bf BCS} is more efficient than {\bf BCS$_g$}. Hence, in the experiments showed in the next section, we implemented {\bf BCS}
and compared it with other methods.
\end{remark}

\subsection{Complexity Comparison}
We compare the complexity of {\bf BCS} with those of the exist CS algorithms. To the best of the authors' knowledge, the unique result for the complexity of the whole process of a CS algorithm is the complexity bound of {\bf TDCS}$_2$, which is $O(2^{\log_2(m)n})$ \citep{cs_fq}. It is easy to see that:
\begin{enumerate}[1)]
  \item when $2\le d < \log_2(m)$, {\bf BCS}  is better than {\bf TDCS}$_2$;

  \item when $d\ge  \log_2(m)$, {\bf BCS} is worse than  {\bf TDCS}$_2$.
\end{enumerate}


Now we compare the complexity of {\bf BCS} with those of other kinds of algorithms for different degree ranges.

\begin{enumerate}[1)]

\item $d=2$:  As mentioned before,  without any side conditions, the complexity  bound of {\bf BCS}  in the worst case is $O(n(m+n)2^{n})$, and this bound is worse than $4\log_2(n)2^n$, which is the complexity bound of the fast exhaustive search method proposed in \citep{fast_search}. For general systems, it is not clear that whether the complexity bound of {\bf BCS} can be improved. However, for other algorithms, its asymptotic complexity can be lower than $O(2^n)$. In this case, the best existing result is  $O(2^{0.841n})$ when $m=n$ \citep{bardet_comp}.

\item $d=3$:  To the best of authors' knowledge, it seems that there are few results about the complexity of solving Boolean polynomial systems when $d>2$. Only some results about the complexity of Gr\"obner basis algorithms are presented in \citep{bardet1, bardet_thesis, bardet_mega}. In \citep{bardet1, bardet_thesis}, the authors show that for solving semi-regular systems with $m=n$,  the degree of regularity $D_{reg}$ is equal to $0.15n+1.35n^{1/3}-1.42+O(\frac{1}{n^{1/3}})$, hence the complexity of the F5 algorithm is $O({n \choose{ D_{reg}}}^\omega)$, where $2< \omega < 3$ is the linear algebra constant. This value is about $O(2^{0.61\omega n})$, and is smaller than $2^{(3-1)n}$, which is the exponential part of the complexity of {\bf BCS}. Therefore, the asymptotic complexity of the F5 algorithm is better than that of {\bf BCS} for semi-regular systems. When considering the input systems without side conditions, the only bound we can know about $D_{reg}$ is $D_{reg}\le n+1$. In this case, the complexity of the F5 algorithms is $O(2^{\omega n})$, hence {\bf BCS} is sightly better.

\item $d\ge 4$:  For the semi-regular systems with $m=n$, when $4\le d \le 7$, the values of $D_{reg}$ are presented in \citep{bardet1, bardet_thesis}:
\begin{itemize}

\item $d=4$, $D_{reg} = 0.20n+1.60n^{1/3}-1.27+O(\frac{1}{n^{1/3}})$;

\item $d=5$, $D_{reg} = 0.24n+1.79n^{1/3}-1.11+O(\frac{1}{n^{1/3}})$;

\item $d=6$, $D_{reg} = 0.26n+1.95n^{1/3}-0.94+O(\frac{1}{n^{1/3}})$;

\item $d=7$, $D_{reg} = 0.28n+2.09n^{1/3}-0.78+O(\frac{1}{n^{1/3}})$.

\end{itemize}

We can check that in these cases, $O({n \choose{ D_{reg}}}^\omega)$ is much less than $O(2^{(d-1)n})$, and $\log({n \choose{ D_{reg}}}^\omega)$ increases much slower than $(d-1)n$ when $d$ increases. For example, when $d=7$,  $\log({n \choose{ D_{reg}}}^\omega)$ is about $0.86\omega n$, which is much less than $(7-1)n=6n$.
Hence  the asymptotic complexity of the F5 algorithm is much better than that of {\bf BCS} for semi-regular systems.
Moreover, when considering the input systems without side conditions,  similarly as the case $d=3$, we also have $D_{reg} \le n+1$, hence the complexity of  the F5 algorithm is bounded by $O(2^{\omega n})<O(2^{3n})$. In comparison, the exponential part of the complexity of {\bf BCS} is $O(2^{(d-1)n})\ge O(2^{3n})$, thus {\bf BCS} is worse than F5 algorithm.

\end{enumerate}

\noindent In all, the above comparison implies that when $d$ is small, {\bf BCS} is much more efficient then existing CS algorithms and  may be comparable with other algorithms, and this is coherent with our experimental observations.

Actually, in a lot of cases, the depth of the zero decomposition tree can be much smaller than $(d-1)n$.
\begin{itemize}
\item For example, in practical computation, after \simplify or \addreduce we may obtain some new linear polynomials or some lower degree monic polynomials, hence $\d_0, \d_i$ may decrease, then $Sum(\T)$ decreases much faster. This always happens when the input systems are sparse or have some algebraic structure. In Section \ref{sec_experiment}, we will show that for sparse polynomial systems even when $d$ is big $(d>4)$, , {\bf BCS} is still very efficient.

\item Another example is the case that the zero set of the input system is large, which implies the dimensions of monic triangular sets corresponding to the solutions are big. Suppose we obtain a monic triangular set with dimension $k$ from one solving branch. This means that after we deal with this solving branch, $k$ entries of $\T$ will be $d$, thus $Sum(\T)\ge dk+n+1-k$. Therefore, the depth of this branch is not larger than $(n+1)d-dk-(n+1-k)=(n+1-k)(d-1)$, and this value is much less than $n(d-1)$ when $k$ is big \footnote {This explains why  {\bf BCS} is much efficient than other algorithms for solving the {\em Matrix}  problems, which have large number of solutions,  showed in the Section \ref{sec_experiment}.}.
\end{itemize}

\section{Experimental results}\label{sec_experiment}

In this section we present some experimental results about {\bf BCS}. We have implemented {\bf BCS} with the C language and the CUDD package (http://vlsi.colorado.edu/~fabio/CUDD) by which the Boolean polynomials are stored as zero-suppressed binary decision diagrams(ZDDs) \citep{zdd, polybori}.
In our implementation, the \fchoose function is originated from $\fchoose_2$. That is for a polynomial $P=Ix_c+U$, we define an index
$$(\tdeg(I),term(I),term(U), \cls(I))$$
where $term(P)$ is the number of monomials in $P$, then choose the polynomial with  the smallest index w.r.t. the lexicographical order. The executable file of our implementation is available at {\em https://github.com/hzy-cas/BCS}.
Our experiments were done on a Macbook Pro with a Intel  i7 2.7 GHz CPU (only one core is used), 8G memory, and  Mac OS X.
We compared {\bf BCS} with the following four methods  by solving the same polynomial system on the same platform.
\begin{enumerate}[(1)]

\item The modified {\bf MFCS} algorithm introduced in Section \ref{sec_algorithm}, and used in the experiments of \citep{cs_fq}, which is available at {\em https://github.com/hzy-cas/MFCS}. We denoted it by {\bf MFCS$_1$}.

\item  The Gr\"obner basis routine over Boolean polynomial ring in Magma V2.20-3 w.r.t. graded reverse lexicographic order, denoted by {\bf BGB}. As mentioned in the handbook of Magma, since V2.15, computing the Gr\"obner bases of an ideal in the Boolean polynomial ring is available, and this routine exploits the properties of Boolean polynomial ring to accelerate the computation.

\item The Gr\"obner basis routine over Boolean polynomial ring in SAGE V8.7, denoted by {\bf Polybori}. This routine is implemented by the {\bf Polybori} library, which is designed for solving the problems of Boolean polynomials and uses ZDDs  as its data structure \citep{polybori}. Since {\bf Polybori} have good performance in computing the Gr\"obner basis w.r.t. the lexicographic order, in the experiments, we recorded two groups of data by using the graded reverse lexicographic order and the lexicographic order respectively.

\item {\bf Cryptominisat} V5.6.8, a SAT-solver which is very efficient for solving SAT problems converted from Boolean polynomial systems hence is a widely used for solving Boolean PoSSo problems \citep{cryptominisat}.
To convert Boolean PoSSo problems to SAT problems, we used the ANF to CNF converter in SAGE V8.7,  which applied a lot of techniques to efficiently convert an ANF to CNF, and recorded the time cost of converting.
 Note that in our experiments, we wanted to achieve all the solutions of the input systems, therefore we used the parameter ``maxsol" such that the solver can output all the solutions.

\end{enumerate}

\noindent
In our experiments, we solved several groups of Boolean polynomial systems which are generated
from some typical algebraic cryptanalysis and reasoning problems. We introduce these problems specifically.

\begin{enumerate}[1)]

\item {\em Present}: a polynomial system originated from the key recovery problem of  the block cipher Present with one pair of known plaintext and ciphertext \citep{present}. Here we consider a reduced version of Present which has only 5 rounds. By setting the 80-bit key as variables $\{x_0, x_1,\ldots, x_{79}\}$ and  adding some internal variables used to simplify the structure of the systems, we generate a  system with $356$ variables, $1785$ quadratic polynomials. Then we randomly guessed the value of variables $\{x_0, x_1, \ldots, x_{47}\}$, and obtain the input system of our experiment.

\item {\em Serpent}:  a polynomial system originated from the problem of recovering the initial key from the 2-round key schedule of the block cipher Serpent. This problem is the basic problem of the cold boot key recovery problem of Serpent \citep{cold_carlos, max_posso_huang}. The system has $128$ variables which corresponding to the 128-bit initial key, $256$ polynomials and degree $3$.

\item {\em MayaSbox}: a polynomial system originated from the problem of recovering a secret 4-bit Sbox, $S: \F_2^4\rightarrow \F_2^4$, of the block cipher Maya from its input and output difference\citep{maya0, maya}.
The variables of this system are corresponding to the different bits of 16 bytes output $S(0000), S(0001), \ldots, S(1111)$, hence the system has $16\times 4=64$ variables. There are $304$
polynomials in this system. $64$ of them are quadratic polynomials representing the input and output differences, and $240$ of them are polynomials with degree 4, which represent the bijection property of the Sbox.

\item {\em Canfil}:  a  polynomial system originated from the stream cipher based on a linear feedback shift register (LFSR) and a filter function \citep{fa1, cs_fq}, and has the form $$\{ f(x_1,x_2,\ldots,x_n), f(L(x_0,x_1,\ldots,x_{n-1})), \ldots, f(L^{m-1}(x_0,x_1,\ldots,x_{n-1}))\}.$$
\begin{itemize}

\item {\em Canfil2}: $n=64$, $m=68$, $L = L_1$, $L_1(x_0,x_1,\ldots,x_{63})=(x_1,x_2,\ldots,x_{63}, x_{63}+x_{59}+x_{46}+x_{45}+x_{36}+x_{30}+x_{24}+x_{18}+x_{14}+x_{11}+x_1+x_0)$,
$f(x_0,x_1,\ldots,x_{63}) = x_5x_{14}+x_0x_{11}+(x_0x_5+1)x_7$.

\item {\em Canfil3}: $n=64$, $m=68$, $L=L_1$,
$f(x_0,x_1,\ldots,x_{63}) = (x_5x_7x_{11}+x_7+1)x_{14}+(x_5+1)x_{11}+x_0x_5x_7,$.

\item {\em Canfil4}: $n=64$, $m=68$, $L=L_1$,
$f(x_0,x_1,\ldots,x_{63}) = x_{0}x_{11}x_{14}+(x_0+1)x_5x_7+x_0$.

\item {\em Canfil5}: $n=64$, $m=68$, $L=L_1$,
$f(x_0,x_1,\ldots,x_{63}) = x_5x_7x_{11}x_{14}+x_5x_7+x_0$.

\item {\em Canfil6}: $n=64$, $m=68$, $L=L_1$,
$f(x_0,x_1,\ldots,x_{63}) =x_0x_5x_7x_{14}+x_{11}+x_5x_7,$.

\item {\em Canfil7}: $n=64$, $m=68$, $L=L_1$,
$f(x_0,x_1,\ldots,x_{63}) =x_5x_7x_{14}+x_5x_7x_{11}+(x_0x_5+1)x_7+x_5+x_0$.

\item {\em Canfil8}: $n=40$, $m=60$, $L(x_0,x_1,\ldots,x_{39})=(x_1,x_2,\ldots, x_{39}, x_{37}+x_{34}+x_{21}+x_{11}+x_{5}+x_0)$,
$f(x_0,x_1,\ldots,x_{39}) = (x_{25}+x_6x_{11})x_{31}+x_{25}+(x_{11}+1)x_{18}+x_0x_6x_{11}+x_0x_6$.

\end{itemize}

\item {\em Biviuma}: a polynomial system originated from the problem of recovering the internal states of stream cipher Bivium-A\citep{bivium, bivium_faugere}, which is a reduced version of  stream cipher Trivium. We set the 177-bit internal states as variables, and add two internal variables at each clock of the cipher. Then by 400-bit keystream,  we  generate a polynomial system with $977$ variables and $1062$ polynomials from 400-bit keystream. $662$ of these polynomials are quadratic, and others are linear.

\item {\em Biviumb}: a polynomial system originated from the problem of recovering the internal states of the stream cipher Bivium-B\citep{bivium, bivium_gb}, which is a reduced version of  stream cipher Trivium. By setting the 177-bit internal states as variables, and using 160-bit keystream, we can generate a polynomial system with 177 variables. This system contains $12$ polynomials with degree $3$, $82$ polynomials with degree $2$, and $66$ polynomials with degree $1$.  Since this system cannot be directly solved by any method in reasonable time, in the experiments, we guessed $33$ variables as the strategy proposed in \citep{cs_africa} to simplify the system.

\item {\em Matrix}: a polynomial system originated from the Boolean matrix multiplication problem proposed by Stephen Cook in his invited talk
at SAT 2004 \citep{sat-cook,book-cook, cs_fq}.
The problem is that given two $k\times k$ Boolean matrices $A$ and $B$, prove $BA=I$ from $AB=I$ by reasoning. By setting the entries of $A$ and $B$ to be $2k^2$ distinct
variables,  we can obtain $k^2$ quadratic polynomials from $AB=I$.
Then the reasoning problem is equivalent to computing the Gr\"obner basis or the zero decomposition of these polynomials, then checking whether the polynomials generated by $BA =I$ can be reduced to 0 by the Gr\"obner basis or by every triangular set in the zero decomposition.
In the following tables,  {\em Matrix3}, {\em Matrix4}, {\em Matrix5}, {\em Matrix6} are the polynomial systems corresponding to the problems with order $3,4,5,6$ respectively.
Note that, since the number of solutions for $AB=I$ is very huge, evidently a SAT-solver cannot output so many solutions in reasonable time.  Therefore, for a SAT-solver, a better way to prove $BA=I$ from $AB=I$ is checking whether the corresponding negative proposition is true. For this purpose, we generate the polynomial system {\em Matrix-neg} introduced below.

\item{\em Matrix-neg}:  A polynomial system corresponding to the negative proposition of the above matrix multiplication problem. Precisely, we generated a polynomial system consists of  polynomials corresponding to  $AB=I$, and one polynomial corresponding to $(BA)_{11}=0$. Here $(BA)_{11}$ is the entry in the first row and first column of $BA$.  It is obvious that this polynomial system has no solution.  When the orders of $A$ and $B$ are $k$, this polynomial system has $2k^2$ variables and $k^2+1$ quadratic polynomials. In the following tables, {\em Matrix3-neg},  {\em Matrix4-neg}, {\em Matrix5-neg}, {\em Matrix6-neg} are the polynomial systems corresponding to the problems with order $3,4,5,6$ respectively.

\end{enumerate}

Besides these polynomial systems generated from cryptanalysis and reasoning, we randomly generated some sparse and dense  polynomial systems with different $n$ and different degrees, then solved them in our experiments. Here we set $m=n$, and in this case,  we found most of these random generated polynomial systems have 0-3 solutions.
\begin{itemize}
\item We generate a sparse polynomial systems with degree $d$ by the following way. For each polynomial, we set the number of its monomials with degree $d_0$ to be $n/2$, where $d_0=2,\ldots,d$, and randomly choose each monomial. Then we randomly generated the constant term of this polynomial.  At last, we will obtain a sparse inhomogenous polynomial with $n$ variables, $nd/2$ non-constant terms and total degree $d$. In the following tables,  we denote such  polynomial systems with $n$ variables and degree $d$  by    {\em RandSparse${(n,d)}$}.

\item To generate a dense polynomial with degree $d$, for each monomial with degree not bigger than $d$, we randomly generate the number $0$ or $1$ with probability 1/2, and if we get $1$, then we let this monomial be in the polynomial. By this way, the expectation number of the monomials in this polynomial will be $\frac{1}{2} \sum_{i=0}^{d}{n \choose i}$.  In the following tables,  we denote such polynomial systems with $n$ variables and degree $d$  by {\em RandDense${(n,d)}$}.

\end{itemize}

Specific instances of these polynomial systems can be found in the {\em benchmarks} directory of the implementation of {\bf BCS} at {\em https://github.com/hzy-cas/BCS}. In the following table, the time costs of solving these systems by different methods are presented.
Note that  the input polynomial systems of {\em Maxtrix} and {\em Matrix-neg} problems are fixed, while other polynomial systems can be generated by random parameters. Hence in our experiments, except {\em Maxtrix} and {\em Matrix-neg} problems, for each other problem, we generated 10 different instances, and the timings presented in these tables are the average time of solving ten instances.

In Table \ref{tb_para}, we show the basic parameters of the polynomial systems generated from cryptanalysis and reasoning. Here, $n$ is the number of variables and $m$ is the number of polynomials. In the column ``degree", since some systems have polynomials with different degrees, for each degree we wrote down the number of polynomials in the brackets.  Table \ref{tb_time},  \ref{tb_random_sparse}, \ref{tb_random_dense} present the timings, which are all given in seconds. In these tables, ``$\#$" means crashed, and  ``$*$" means running over 2 hours without output. Note that, as in \citep{satconversion} the timings of {\bf Cryptominisat} in these tables are the sums of the time of converting and the time of solving. Moreover, since  for solving these random generated systems, {\bf MFCS}$_1$ is always worse than {\bf BCS}, and {\bf Polybori} is always worse than {\bf BGB}, we only list the timings of {\bf BCS}, {\bf BGB} and {\bf Cryptominisat} in Table  \ref{tb_random_sparse} and Table \ref{tb_random_dense}, in order to show the evolution of the timings  for these three kinds of methods with different $n$ and $d$.

\begin{table}[!ht]\centering
\caption{The basic parameters of the input polynomial systems}
\vspace{5pt}
{\small
\begin{tabular}{c c c c}\hline	\label{tb_para}
{ \em Benchmarks} \SPC & n & m & degree  \\ \hline
 {\em MayaSbox} & 64 & 304 &  2(64), 4(240)   \\
 {\em Serpent} & 128 & 256 & 3(256) \\
  {\em BiviumB}   & 177 & 193 &  1(99), 2(82), 3(12)  \\
 {\em Present}     & 356  & 1833 & 1({\footnotesize 48}), 2(1785)  \\
 {\em BiviumA}  & 977 & 1062 & 1(400), 2(662)   \\ \hline

 {\em Canfil2}  & 64 & 68 & 3(68)       \\
  {\em Canfil3} & 64 & 68 & 3(68)      \\
  {\em Canfil4} & 64 & 68 & 3(68)      \\
  {\em Canfil5} & 64 & 68 & 4(68)   \\
  {\em Canfil6} & 64 & 68 & 4(68)      \\
 {\em Canfil7} & 64 & 68 & 3(68)     \\
 {\em Canfil8}  & 40 & 60 & 3(60)   \\  \hline

 {\em Matrix3} &18 & 9 & 2(9) \\
  {\em Matrix3-neg} &18 & 10 & 2(10) \\
 {\em Matrix4} & 32 & 16 & 2(16) \\
  {\em Matrix4-neg} & 32 & 17 & 2(17) \\
 {\em Matrix5} &  50 & 25 & 2(25)   \\
  {\em Matrix5-neg} &  50 & 26 & 2(26)   \\
 {\em Matrix6}   & 72 & 36  & 2(36)   \\
 {\em Matrix6-neg}   & 72 & 37  & 2(37)   \\
 \hline
\end{tabular}
}
\end{table}

%

{\footnotesize
\begin{table}[!ht]\centering
\caption{Timings for solving polynomial systems from cryptanalysis and reasoning }
\vspace{5pt}
{\small
\begin{tabular}{l r r r r r r r}\hline	\label{tb_time}
 Benchmarks  &{\bf BCS} & {\bf MFCS$_1$} & {\bf BGB}& {\bf Polybori}& {\bf Polybori}   & {\bf Cryptominisat}  \\
    &  & &  [grevlex] &  [lex] & [ grevlex] & \\  \hline
\vspace{2pt}
 {\em MayaSbox}  &0.10  &  0.24 &  0.90 & 95.57 & 93.27  &  35.19
 \\\vspace{2pt}
 {\em Serpent}  &0.40    &  0.47   &  110.79 & 176.00   &  171.80 &  292.19\\\vspace{2pt}
  {\em BiviumB}   &14.17 & 18.45  &  61.93  & $*$ & 358.98  & 5512.55 \\\vspace{2pt}
 {\em Present}    &5.97  &  14.86 & 511.49  & 63.00 &95.00 &  21.70 \\ \vspace{2pt}
 {\em BiviumA} &13.52 &  37.36 & 2106.29  & $*$   &  $\#$ &76.27 \\ \hline \vspace{2pt}
 {\em Canfil2}  &15.36 & 16.88  &  $*$   &     $\#$ &$\#$  & 150.07    \\\vspace{2pt}
  {\em Canfil3} &41.83 & 50.13  &  891.63  &     $\#$ & $\#$  & 2269.83     \\\vspace{2pt}
  {\em Canfil4}  &1.01 & 1.95  &  107.75   &     15.50  & 13.90 & 149.63       \\\vspace{2pt}
  {\em Canfil5}  &28.99 & 34.05  &  356.25  &    277.77 & 215.32 & 2778.22
       \\\vspace{2pt}
  {\em Canfil6} &6.69	 & 	10.25& 208.93  &     22.7 & 22.02 & 2062.02
      \\ \vspace{2pt}
 {\em Canfil7} &6.07 & 6.84  & 2721.54   &     $\#$ & $\#$   &   159.55\\ \vspace{2pt}
 {\em Canfil8}     &392.10  & $*$  & $*$  &  $\#$ & $\#$   & 690.64 \\ \hline\vspace{2pt}
 {\em Matrix3}   &0.002  & 0.002
 &0.66 & 3.21 & 1.23  & 0.53 \\ \vspace{2pt}
   {\em Matrix3-neg} &  0.002
  &  0.002 &0.02 & 0.57 & 0.52 & 0.55   \\\vspace{2pt}
 {\em Matrix4}   & 0.02  & 0.03 & 2436.04 & $\#$ & $\#$  &  41.46\\ \vspace{2pt}
  {\em Matrix4-neg} & 0.03 &  0.03 & 733.38 &465.51& 450.48  & 0.48\\\vspace{2pt}
 {\em Matrix5}   &0.60  &  
 14.29
 &$*$ & $\#$ & $\#$  & $*$   \\\vspace{2pt}
 {\em Matrix5-neg} &   2.57  &  10.25  &1645.76 & $\#$  &  $\#$ & 148.40 \\\vspace{2pt}
 {\em Matrix6}   & 85.44  &  
 $*$& $*$  & $\#$ & $\#$  & $*$\\\vspace{2pt}
 {\em Matrix6-neg}   & 418.88  &  2716.53
& $*$  & $\#$ &  $\#$ & $*$ \\
 \hline
\end{tabular}
}
\end{table}

\begin{table}\centering
\caption{Timings for solving random sparse polynomial systems with $m =n$}
\vspace{5pt}
{\small
\begin{tabular}{l  l r r r r r r}\hline	\label{tb_random_sparse}
{ Benchmarks $(n, d)$}& {\bf BCS} &{\bf BGB}&  {\bf Cryptominisat}  \\
\hline
\vspace{2pt}
 {\em RandSparse$(22, 2)$} &  0.81  & 9.66  & 3.37 \\ \vspace{2pt}
 {\em RandSparse$(22,3)$} &   3.80  & $\#$ & 10.42 \\\vspace{2pt}
 {\em RandSparse$(22,4)$} &  11.72 & $\#$  &  20.35\\\vspace{2pt}
 {\em RandSparse$(22,5)$} &   21.69   & $\#$ & 31.46\\\vspace{2pt}
 {\em RandSparse$(22,6)$}  & 35.48 &$\#$  &  55.24  \\\vspace{2pt}
 {\em RandSparse$(22,7)$} & 43.25 &  $\#$ &  58.70     \\
 \hline

  {\em RandSparse$(26, 2)$} &   9.51   & 202.63  &   19.65  \\\vspace{2pt}
  {\em RandSparse$(26, 3)$} &    68.94    &  $\#$& 117.98\\\vspace{2pt}
  {\em RandSparse$(26, 4)$} &    162.09 &  $\#$ & 218.79
       \\\vspace{2pt}
  {\em RandSparse$(26, 5)$} &  356.63 & $\#$ &389.17
      \\ \vspace{2pt}
 {\em RandSparse$(26, 6)$}&    552.36 & $\#$ & 599.02  \\ \vspace{2pt}
 {\em RandSparse$(26, 7)$}   &     723.15   & $\#$ & 840.91 \\
 \hline

   {\em RandSparse$(30, 2)$}      &112.91&  $*$& 206.68    \\\vspace{2pt}
  {\em RandSparse$(30, 3)$}    &1198.79 & $\#$ &3837.57   \\\vspace{2pt}
  {\em RandSparse$(30, 4)$}   &3185.50 & $\#$ & $*$
       \\\vspace{2pt}
  {\em RandSparse$(30, 5)$}   &5356.85 & $\#$ & $*$
      \\ \vspace{2pt}
 {\em RandSparse$(30, 6)$}    &$*$  &  $\#$ &  $*$\\ \vspace{2pt}
 {\em RandSparse$(30, 7)$}      & $*$  &    $\#$ & $*$ \\ \hline

    {\em RandSparse$(34, 2)$}     &1147.51&  $*$ & $*$  \\\vspace{2pt}
  {\em RandSparse$(34, 3)$}    &$*$ & $\#$ &   $*$   \\\vspace{2pt}
  {\em RandSparse$(34, 4)$}     &$*$ & $\#$& $*$
       \\ \hline
\end{tabular}
}
\end{table}

\begin{table}\centering
\caption{Timings for solving random dense polynomial systems with $m=n$}
\vspace{5pt}
{\small
\begin{tabular}{l r r r r r r}\hline	\label{tb_random_dense}
Benchmarks $(n,d)$  &{\bf BCS}& {\bf BGB}&{\bf Cryptominisat}  \\\vspace{2pt}
{\em RandDense$(18,2)$} &  1.32  &    0.44 & 14.33  \\\vspace{2pt}
{\em RandDense$(18,3)$} &   64.77     &   41.99  &   84.75 \\\vspace{2pt}
{\em RandDense$(18,4)$} & 757.13  &883.42 & 363.68  \\\vspace{2pt}
{\em RandDense$(18,5)$} &  $*$ &  $*$ & 1160.61     \\
{\em RandDense$(18,6)$} &  $*$ & $ *$ &$*$    \\ \hline

{\em RandDense$(20,2)$} &  4.95 & 1.03 & 53.59 \\\vspace{2pt}
{\em RandDense$(20,3)$} &   429.93     & 2080.61     & 420.13 \\\vspace{2pt}
{\em RandDense$(20,4)$} & 7162.80 &$ *$   &2458.26   \\\vspace{2pt}
{\em RandDense$(20,5)$} &  $*$ & $*$ &$*$     \\\hline

{\em RandDense$(22,2)$} &   21.55 &  9.77 &    193.26\\\vspace{2pt}
{\em RandDense$(22,3)$} &   2755.51  &   $\#$ & 3208.67\\\vspace{2pt}
{\em RandDense$(22,4)$} &   $*$  &   $\#$  &  $*$\\\hline

{\em RandDense$(24,2)$} &  94.23  &  41.38 &   1296.9   \\\vspace{2pt}
{\em RandDense$(24,3)$} &  $*$  &$\#$&  $*$  \\\vspace{2pt}
{\em RandDense$(24,4)$} & $*$  & $\#$&  $*$   \\\hline

  {\em RandDense$(26, 2)$} &  379.24   &   249.76 &  $*$  \\\vspace{2pt}
  {\em RandDense$(26, 3)$} &   $*$  & $\#$&  $*$   \\\hline

   {\em RandDense$(28, 2)$} & 1587.26 & $*$&    $*$\\\vspace{2pt}
  {\em RandDense$(28, 3)$} &  $*$ &   $\#$ &   $*$     \\\hline

   {\em RandDense$(30, 2)$} &    7038.65 & $*$ &  $*$ \\ \vspace{2pt}
  {\em RandDense$(30, 3)$} &   $*$  & $\#$ & $*$  \\ \hline
\end{tabular}
}
\end{table}
}

From Table \ref{tb_time}, we can observe that {\bf BCS} is the most efficient algorithm for solving any of these systems generated from cryptanalysis and reasoning.
We think that one reason of {\bf BCS} being so efficient is that these polynomial systems have some block triangular structure, which means the classes of the polynomials can be divided into different sets. Moreover, most of the polynomials in these systems are sparse.  Therefore, the decreasing of $Sum(T)$  for {\bf BCS} is very fast.

\

From Table \ref{tb_random_sparse} and Table \ref{tb_random_dense}, we have the following observations.
\begin{itemize}
\item  For random sparse systems, we can see that {\bf BCS} is the most efficient algorithm.
Note that,  the influence of the degree of the input systems to the timings is much weaker than we expected. For $d>3$, when $d$ increases by 1, the timing of {\bf BCS} increases less than double. Moreover, when $n$ increases by $4$, the timing increases about $2^{0.9\times 4}$ times. All these observations show that {\bf BCS} can well use the sparse property, hence have a much lower practical complexity.
This raises a question: can we  achieve a lower asymptotic complexity bound about {\bf BCS} when the input polynomial systems are sparse?

\item For random dense systems, when $d=2$,  {\bf BCS} is comparable with other algorithms, and {\bf BGB} is the most efficient one.  Moreover,  when $n$ increases by 2, the timing of {\bf BCS} increases about $2^2$ times, and this is consistent with our asymptotic complexity for quadratic systems.
When $d=3$,  {\bf BCS} is comparable with {\bf Cryptominisat}, and {\bf BGB} didn't work well for $n\ge 20$, because the degree of regularity is too high.  When $d\ge 4$,  {\bf Cryptominisat} becomes the most efficient algorithm, and this is coherent with our prediction. Since for random dense polynomial systems with high degree,  few algebraic properties can be exploited,  the methods based on the idea of searching the values of variables will be more efficient. Actually,  fast exhaustive search algorithms, for example, libFES\citep{fast_search},  are much more efficient than all these algorithms for solving random dense polynomial systems, but they are not capable for solving most of the problems in Table \ref{tb_time}, since $n$  is too big for these problems.  Note that, for these systems, when $d$ increases by 1, the timing of {\bf BCS} increases about $2^6$ to  $2^7$ times, which is much less than $2^n$ times. It seems that our asymptotic complexity bound can be improved even for random dense systems.
\end{itemize}

Now, we show why {\bf BCS} is more efficient than other CS algortihms.
Compared to {\bf MFCS$_1$} and {\bf MFCS}, the mainly advantage of {\bf BCS} is that the number of branches is smaller. This can be well explained by analyzing the change of $Sum(T)$ after zero decomposition.

\begin{itemize}
\item In {\bf MFCS}, $Sum(T)$ decreases slowly. For example,  suppose there are $k$ polynomial $\{f_1, f_2,\ldots, f_k\}$ with the highest class and highest degree, and $\tdeg(\initial(f_1)) = \tdeg(\initial(f_2)) = \cdots =
\tdeg(\initial(f_k))$. Suppose we have executed zero decomposition w.r.t. $f_1$.  Then, $Sum(T)$ will not change after the next $k-1$ times of decomposition were finished. The reason is that we will choose $f_2, \ldots, f_k$ to do zero decomposition,  and $\d_0, \d_1, \ldots, \d_n$ will not change.
As mentioned in Section \ref{sec_complexity}, in the worst case, the depth of the zero decomposition tree of {\bf MFCS} can reach the bound proposed in Proposition \ref{th_complexity1}.

\item In {\bf MFCS$_1$}, $Sum(\T)$ decrease faster than in {\bf MFCS}. For example,  if $P$ is the polynomial with shortest initial, then after choosing $P$ to do zero decomposition, in a lot of cases, $\initial(P)$ will also be the polynomial with shortest initial, hence $\d_0$ will decrease.
However, there are still some cases that $Sum(\T)$ doesn't decrease after zero decomposition.
For example, if $\initial(P)$ is monic, and $\d_{c} < \tdeg(\initial(P)), c=\cls(P)$, since we don't compute add-remainder and {\addreduce} will be executed until all polynomials are monic, then in the following process,  we may choose a polynomial $Q$ with $\tdeg(Q)\ge \tdeg(P)$ to do zero decomposition. It means that $\d_{c}$ will not change and $\d_0$ will not decrease, hence $Sum(\T)$ will not decrease.

\item In {\bf BCS}, from Section \ref{sec_complexity}, we know that if the choose function is $\fchoose_1$(or $\fchoose_2$), $Sum(\T)$(or $Sum(\T')$) strictly decreases after zero-decomposition, which means we don't have ``useless" decomposition in {\bf BCS}, hence the number of branches is smallest.

\end{itemize}

\section{Conclusion}
In this paper, we present an improved characteristic set algorithm {\bf BCS} to solve Boolean polynomial systems. This algorithm is based on the idea of eliminating variables by addition and some important techniques.  We introduce the idea of the zero decomposition tree, by which we convert the problem of estimating the complexity of {\bf BCS} into estimating the complexity of solving one branch and the depth of the tree.
We define an index vector about the lowest degree of the non-monic polynomials and monic polynomials with different classes, and give some bounds about the depth of the zero decomposition tree by analyzing the variation of $Sum(\T)$, which is the sum of the entries of this index vector. In this way, we obtain some bit-size complexity bounds of {\bf BCS}, which are lower than those of previous characteristic set algorithms. Moreover, by $Sum(\T)$, we illustrate how the techniques we used in {\bf BCS} effect the depth of the zero decomposition tree.
Furthermore, we test {\bf BCS} by solving some random generated polynomial systems and  some polynomial systems generated from cryptanalysis and reasoning problems. Experimental results show that {\bf BCS} is more efficient than the previous characteristic set algorithms, and comparable with other efficient algorithms.
It is our future work to see whether we can obtain some lower complexity bounds about the algorithm when the input systems are sparse.

\end{document}